\documentclass[twocolumn]{article}
\usepackage{graphicx}
\begin{document}
\title{On the time evolution of giant radio galaxies}
\author{J. Machalski,$^1$\thanks{Email:
 machalsk@oa.uj.edu.pl} K.T. Chy\.{z}y,$^1$ M. Jamrozy$^2$\\
$^1$Astronomical Observatory, Jagellonian University,
ul. Orla 171, PL-30244 Cracow, Poland\\
$^2$Radioastronomisches Institut der Universit\"{a}t Bonn,
Auf dem H\"{u}gel 71, D-53121 Bonn, Germany}

\maketitle
\begin{abstract}
A time evolution of {\sl giant} lobe-dominated radio galaxies
(with projected linear size $D>1$ Mpc for
$H_{0}$=50 km\,s$^{-1}$Mpc$^{-1}$ and $q_{0}$=0.5) is analysed on the basis
of dynamical evolution of the entire FRII-type population. Two
basic physical parameters, namely the jet power $Q_{0}$ and central density of
the galaxy nucleus $\rho_{0}$ are derived for a sample of {\sl giants} with
synchrotron ages reliably determined, and compared with the
relevant parameters in a comparison sample of normal-size sources consisting of
3C, B2, and other sources. Having apparent luminosity (power) $P$ and linear
size $D$ of each sample source, $Q_{0}$ and $\rho_{0}$ are fitted using the
dynamical model of Kaiser et al. (1997) but modified by allowing an evolution
of the cocoon axial ratio with time as suggested by Blundell et al. (1999). 
In a result, we found that:
(i) there is not a unique factor governing the source size; {\sl giants} are 
old sources with high enough jet power evolved in a relatively 
low-density environment. The
size is dependent, in order of decreasing partial correlation
coefficients, on age; then on $Q_{0}$; next on $\rho_{0}$, (ii) a number of the
sample sources, having similar $Q_{0}$ and $\rho_{0}$ but different ages and
axial ratios (called a `clan'), fit very well the evolutionary 
luminosity--size ($P$--$D$) and energy density--total energy
($u_{\rm eq}$--$E_{\rm tot}$) tracks derived from the model for a `fiducial'
source with $Q_{0}$ and $\rho_{0}$ equal to the means of relevant values
obtained for the `clan' members. Therefore, these sources
can be considered as `the same' source observed at different epochs of its
lifetime, and hence very useful for an observational constraint of evolutionary
models of the source dynamics, 
(iii) in some cases a slow acceleration of the average expansion speed 
of the cocoon along the jet axis seems to be present, and (iv) apparent increase 
of the lowest internal pressure value, observed within the largest sources' cocoon, 
with redshift is obscured by the
intrinsic dependence of their age on redshift, which hinders making definite
conclusions about a cosmological evolution of intergalactic medium (IGM) 
pressure.
\end{abstract}

\section{Introduction}

Extragalactic radio sources, powered by twin jets resulting from nuclear
energy processes in the Active Galactic Nucleus (AGN), exhibit a very large
span of their linear size. The sizes of these powerful sources range from
less than $10^{2}$ pc (GPS: gigahertz-peaked spectrum), through $10^{2}-10^{4}$
pc (CSS: compact steep spectrum), $10^{4}-10^{6}$ pc (normal-size sources), up 
to greater than $10^{6}$ pc $\equiv 1$ Mpc (`giant' radio sources). One of the 
key problems of the evolution of extragalactic sources is whether and how 
different
size sources are related. Is there a single evolutionary scheme governing the
size evolution of radio sources, or do small and large sources evolve in a
different way?

From many years `giant'\footnote{hereafter we use {\sl giant} or {\sl giants}
instead of giant radio source(s)} radio sources are of a special interest for
several reasons. Their very large angular sizes give excellent opportunity for
the study of source physics. They are also very useful to study the density and
evolution of the intergalactic and intracluster environment (cf. Subrahmanyan
\& Saripalli 1993;
Mack et al. 1998), as well as to verify the unification scheme for powerful
radio sources (Barthel 1989; Urry \& Padovani 1995). Finally, they can be used
to constrain dynamical models of the source time evolution (e.g. Kaiser \&
Alexander 1999).  The general questions are: do the largest radio sources reach
their extremal {\sl giant} sizes due to (i) exceptional physical conditions
in the intergalactic medium, (ii) extraordinary intrinsic properties of the AGN,
or simply (iii) because they are extremely old?

To answer these questions, in a number of papers the attempts were undertaken
to recognize eventual other properties but size which may differentiate {\sl
giants} from normal-size sources. The {\sl giant}-source morphologies, energy
density in the lobes and interaction with the intergalactic medium were studied
by Subrahmanyan et al. (1996), who suggested that {\sl giant} radio galaxies may
be located in the lowest density regions, and `may attained their large size as
a result of restarting of their central engines in multiple phases of activity
along roughly similar directions'. Also Mack et al. (1998), after a study of 5
nearby {\sl giant} radio galaxies, argued that those sources are so huge because
of their low-density environment and not because of old ages. A similar
conclusion was drawn by Cotter (1998) after his study of a sample of
high-redshift 7C {\sl giants}. He found those sources to be not old and of similar
kinetic powers of the jet as normal-size 3C sources. But the densities of their
surrounding medium was found much lower than those around most of 3C sources
(Rawlings \& Saunders 1991). Ishwara-Chandra \& Saikia (1999) compiled a sample
of more than 50 known {\sl giant} sources (many of them being of FRI-type 
morphology)
and compared some of their properties with those of a complete sample of 3CR
sources `to investigate the evolution of giant sources, and test their
consistency with the unified scheme for radio galaxies and quasars'. They
concluded that the location of {\sl giants} on the power--linear size ($P-D$)
diagram may suggest that the largest sources have evolved from the smaller
sources. Finally, in the recent extensive study of 26 {\sl giant} radio galaxies 
by Schoenmakers et al. (2000), the authors argued that those galaxies `are both
old sources, in term of their spectral age, and are situated in a relatively
low-density environment, but also that neither of these two properties are
extreme. Therefore, their large size probably results from a combination of
these properties'.

From the above results, it is clear that the phenomenon of the {\sl giant} radio
sources is still open to further research. Therefore, in this paper we analyse
whether observed properties of {\sl giant} radio galaxies can be explained by a 
model of the dynamical evolution of classical double radio sources in cosmic time, 
and what factor (if there is a one) is primarily responsible for the {\sl giant}
size. Two recent analytical models, published by Kaiser et al. (1997)
[hereafter referred to as KDA] and
Blundell et al. (1999), are very convenient for this purpose. The overall
dynamics of a FRII-type source (precisely: its cocoon) in both models is based
on the earlier self-similar model of Kaiser \& Alexander (1997) [hereafter
referred to as KA], in which the linear size is very sensitive to a density of
the source environment and a rate of its evolution with cosmic time. The models
of KDA and Blundell et al. differ
between themselves in predictions of a time evolution of the source luminosity.
In the KDA model the synchrotron radiating particles expand from the leading
head into the cocoon whose pressures are in a constant ratio during the source
lifetime. This ensures the constant adiabatic losses in a self-similar expansion
of the cocoon with constant geometry (axial ratio). In the model of Blundell et
al. the radiating particles drift out of a constant pressure hotspot region into
the cocoon in which pressure decreases throughout the source lifetime. This
causes that both the expansion losses and the source axial ratio steadily
increase with time, inducing a faster decrease of source (cocoon) luminosity.

In this paper,
synchrotron ages in a sample of {\sl giant} sources with FRII-type (Fanaroff \&
Riley 1974) morphology are used to verify the dynamical time evolution of such
sources predicted by the KDA model. It is chosen here for its simplicity
in spite of some objections about its application to large and old sources in
which an internal pressure cannot be always above the external medium pressure,
as required by a self-similar model (e.g. Hardcastle \& Worrall 2000).
Basic physical parameters, namely the jet power $Q_{0}$, central density
of the galaxy nucleus $\rho_{0}$, energy density and pressure in the lobes or
cocoon ($u_{c}$ and $p_{c}$), and total energy of the source $E_{\rm tot}$ are
derived from the model for each of member of the sample to fit its redshift,
radio luminosity, projected size, and axial ratio. Next, these physical
parameters are compared with (1) the relevant parameters derived for normal-size
sources in a comparison sample, and with (2) the parameters taken from
observational data, i.e. the age, equipartition energy density $u_{\rm eq}$,
total energy $U_{\rm eq}$, lobes pressure, etc.,
calculated under `minimum energy' conditions. The observational data used are
given in Sect.~2. The method of a comparative analysis of {\sl giants} and
normal-size sources is outlined in Sect.~3. The application of the dynamical
model is described in Sect.~4, while in Sect.~5  results of the modelling are
presented.
The evolutionary tracks of the sources on the $P$--$D$ and
$u_{\rm eq}-E_{\rm tot}$ planes are derived  and confronted with the
observational data in Sect.~6. In the result, we realize that a modification of
the original KDA model is necessary to provide a satisfactory fits to the data.
Discussion of the obtained results and final conclusions are given in Sect.~7.

\section{Observational data}

Similarly to the approach of Ishwara-Chandra \& Saikia (1999), we have compiled 
a sample 
of 18 {\sl giant} sources and a comparison sample of 49 normal-size sources for
which their spectral age was available from the literature. Both samples are
confined to FRII-type sources only. The spectral ageing data for the
{\sl giants} are taken from the papers of Saripalli et al. (1994), Klein et al.
(1996), Mack et al. (1998), Schoenmakers et al. (1998, 2000), Ishwara-Chandra
\& Saikia (1999), Lara et al. (2000), and Machalski \& Jamrozy (2000). For the
aim of this paper, i.e. for an observational verification of the dynamical time
evolution of {\sl giant} sources predicted by its analytical model, the 
comparison sample has been chosen to comprise high-luminosity (high- and
low-redshift), as well as low-luminosity normal-size sources. The high-redshift
(with $z\geq$0.5) and low-redshift ($z<$0.5) subsamples consist of 3C sources 
taken from the papers of Alexander \& Leahy (1987), Leahy et al. (1989), and Liu 
et al. (1992). All of them have $P_{178}\geq 10^{25}$\,W\,Hz$^{-1}$sr$^{-1}$
(other selection criteria are summarized in Liu et al.). A low-luminosity
subsample comprises FRII-type sources with $P_{1.4}< 10^{24.4}$
W\,Hz$^{-1}$sr$^{-1}$ (corresponding to $P_{178}< 10^{25}$\,W\,Hz$^{-1}$sr$^{-1}$
assuming a mean spectral index of 0.7 between 178 and 1400 MHz). A limited
number of such sources with spectral ages determined have been available from
the papers of Klein et al. (1995) and Parma et al. (1999). An additional
criterion applied in our study was a consistence of ages or expansion velocities 
given in the original papers with spectral ages homogeneously estimated by us 
from a fitted spectrum of the sources' cocoon using the standard ageing analysis. 
The use of the spectral age is justified in Sect.~3.1 where the details of our
analysis are given. The wavelength-dependent 
observational data are taken at $\lambda=21$ cm at which a large variety of radio 
maps are available for all the sources.

The observational data for the sample of {\sl giant}-size and normal-size sources
are given in the Appendix.

\section{Method of analysis}

To start a comparative analysis of {\sl giants} and normal-size sources, we have
to determine a number of physical parameters of these sources which must be
connected to their dynamical evolution. All of them are derived from the
observational data using some theoretical models of a classical double radio
source. In this paper we consider the following parameters: (i) the spectral
age, equipartition magnetic field and total energy density derived from the
standard `ageing analysis' of relativistic particles in a steady-state
synchrotron sources (e.g. Miley
1980; Myers \& Spangler 1985), and (ii) the jet power, density of the
intergalactic medium and its distribution, energy densities of relativistic
particles and magnetic field, pressures at a head of the jet and in the cocoon,
and total energy of a source derived from the analytical KDA model.

\subsection{The spectral age}

A determination of the age of the radio sources is crucial to constrain the
dynamical model of time evolution. A characteristic lifetime of sources can be
estimated from their total energy, $E_{\rm tot}$, determined under the
`minimum energy' consideration and the observed power $dE/dt$. The resultant
lifetime, being rather an upper limit to the age of source, is usually greater
than the synchrotron age of relativistic particles commonly determined from the
spectral-ageing analysis (e.g. Alexander \& Leahy 1987). However, the
time-dependence of various energy losses suffered by the particles cause that
different parts of the lobes or cocoon have different ages. Besides, the
radiation losses (and thus a synchrotron age) depend on a history of the
particle injection, a distribution of the pitch angle, etc. described by the
different synchrotron models: Kardashev--Pacholczyk (KP); Jaffe \& Perola (JP);
or continuous injection (CI), cf. Carilli et al.(1991) for a detailed
description. An attempt to minimize the discrepancies between spectral and
dynamical ages has been undertaken by Kaiser (2000). His 3-dimensional model of
the synchrotron emissivity of the cocoon traces the individual evolution of
parts of the cocoon and provides, according to the author, more accurate
estimate for the age of a source. Its application to the lobes of Cygnus~A gave
a very good fit to their observed surface brightness distribution.

However, since an application of the above model is confined to the sources for
which their lobes are reasonable resolved in direction perpendicular to the jet
axis -- we cannot use it for our statistical approach to the dynamical evolution
of {\sl giant} radio sources and have to base on the standard ageing analysis. 
In order to estimate a spectral age of the sample sources as homogeneously as
possible, we have used a spectrum of the sources' cocoon or their two lobes,
i.e. the spectrum of entire source after a subtraction of evident hot spots 
and
bright cores emission (e.g. the core in 3C236). This spectrum, fitted at least
between 151 MHz and 5 GHz, has been used to estimate a synchrotron age of the
cocoon, $t_{\rm syn}$, with the prescription of Myers \& Spangler (1985) for
the JP model. Applying their plot of spectral index between 1446 and 4885
MHz, as well as that between 151 and 1490 MHz given in Leahy et al. (1989),
we were able to find the best injection spectral index and the break frequency.
The resultant ages were then compared with synchrotron ages of the sources
either given explicitly in the cited papers or calculated by us from the 
published expansion velocities (the reference papers are given in column 9 of 
Table A1). If the former age agreed with the latter (to $\pm$30 per cent), the 
source
is included to the samples. The above procedure was necessary because usually 
the values published in different papers were not derived in a uniform way. On
the other hand side, it assured us that our estimates are not discrepant with
those of other authors. Basing on a commonly accepted assumption about a 
proportionality of the spectral  and dynamical ages, hereafter we assume
$t_{\rm dyn}=2t_{\rm syn}$ (e.g. Lara et al. 2000). This age (marked by $t$[Myr])
is given in column 2 of Table~A2.

\subsection{The equipartition magnetic field and energy density}

Also the equipartition magnetic field strength, $B_{\rm eq}$, and energy density,
$u_{\rm eq}$, in the cocoon or the lobes of the sample sources are homogeneously
calculated using the method outlined by Miley (1980), and compared with their
values given in the published papers (cf. Sect.~2). Total luminosity of the
cocoons has been integrated between 10 MHz and 100 GHz using their fitted radio 
spectrum (cf. Sect.~3.1). The volume of the cocoons is calculated assuming a 
cylindrical geometry with the base diameter resulting from the axial ratio 
determined in the radio maps (cf. the Appendix), and the length $D/\sin\theta$,
where $\theta$ is an orientation angle of the jet axis to the observer's 
line-of-sight. The values of $u_{\rm eq}$ and $B_{\rm eq}$ with their estimated
error, calculated with assumption of the filling factor of unity and equal
distribution of energy between electrons and protons, are given in columns 3
and 4 of Table~A2, respectively.

\section{Application of the analytical model}

Below we summarize the KA and KDA dynamical models,
and precise their application in our analysis. It is assumed in the model
that the radio structure is formed by two jets emanating from the AGN into a
surrounding medium in two opposite directions, then terminating in strong
shocks, and finally inflating the cocoon. A density distribution of the
unperturbed external gas is approximated by a power-law relation
$\rho_{\rm d}=\rho_{0}(d/a_{0})^{-\beta}$, where $d$ is the radial distance from
the core of a source, $\rho_{0}$ is a constant density at the core radius
$a_{0}$, and $\beta$ is a constant exponent in this distribution.

The half of the cocoon is estimated by a cylinder of length $L_{\rm j}=D/2$ and
axial ratio $R_{\rm T}=AR/2$. It expands along the jet axis driven by the hot
spot pressure $p_{\rm h}$ and in the perpendicular direction by the cocoon
pressure $p_{\rm c}$. In the model the rate at which energy is transported
along each jet (of power $Q_{0}$) is constant through the source lifetime.
The model predicts self-similar expansion of the cocoon and gives analytical
formulae for the time evolution of various geometrical and physical parameters,
e.g. the cocoon pressure $p_{\rm c}$, its linear size $L_{\rm j}$ and luminosity
$P_{\nu}$.

However, the pressure ratio ${\cal P}_{\rm hc}\equiv p_{\rm h}/p_{\rm c}=
4R_{\rm T}^{2}$, implied in the original KDA paper, has
later been found to seriously overestimate a value of ${\cal P}_{\rm hc}$
obtained in hydrodynamical simulations by Kaiser \& Alexander (1999). Therefore,
in our modelling procedure we use the empirical formula taken from Kaiser (2000):

\begin{equation}
{\cal P}_{\rm hc}=(2.14-0.52\beta)R_{\rm T}^{2.04-0.25\beta}.
\end{equation}

\noindent
Moreover, taking into account the relation pointed out in Sect.~5.2.3, the
KDA model is modified in our paper by a subjection of
$R_{\rm T}$ on age $t$ and $Q_{0}$.

\subsection{Radio power}

The radio luminosity of the cocoon $P_{\nu}$ is calculated in the model by
splitting up the source into small volume elements and allowing them to evolve
separately. The effects of adiabatic expansion, synchrotron losses, and inverse
Compton scattering on the cosmic microwave background radiation are traced for
these volume elements independently. The total radio emission at a fixed
frequency $\nu$ is then obtained by summing  up the contribution from all such
elements, resulting in an integral over the time [Eq.(16) in KDA].
The integral is not analytically soluable and must be calculated numerically.

The predicted radio luminosities can be used to construct evolutionary tracks
of sources on the $P-D$ diagram. These tracks can be compared with the observed
distribution of sample sources on this diagram, and answer the question whether
{\sl giant} sources evolve in the same way as normal-size sources, or not.

\subsection{Source energetics}

In our approach we neglect thermal particles, hence the overall source dynamics
is governed by the pressure in the cocoon in the form
$p_{\rm c}=(\Gamma_{\rm c}-1)(u_{\rm e}+u_{\rm B})$, where $\Gamma_{\rm c}$ is
the adiabatic index of the cocoon, $u_{\rm e}$ and $u_{\rm B}$ are energy
densities of relativistic particles and magnetic field, respectively. Both
energy densities are a function of the source lifetime $t$. In particular,
$u_{\rm B}(t)\propto B^{2}(t)={\rm const}\,t^{-a}$, where
$a=(\Gamma_{\rm B}/\Gamma_{\rm c})(4+\beta)/(5-\beta)$.

Since the time evolution of the  pressure $p_{\rm c}$ is known from the
self-similar solution, then one can calculate the energy density in the cocoon
at any specific age $t$:

\[u_{\rm c}(t)\equiv u_{\rm e}(t)+u_{\rm B}(t)=p_{\rm c}(t)/(\Gamma_{\rm c}-1),\]

\noindent
and the total source energy: $E_{\rm tot}(t)=u_{\rm c}(t)V_{\rm c}(t)$, where
$V_{\rm c}$ is the cocoon volume attained at the age $t$:

\begin{equation}
V_{\rm c}(t)=2\frac{\pi}{4R_{\rm T}^{2}}[L_{\rm j}(t)]^{3}\propto t^{9/(5-\beta)}.
\end{equation}

\noindent
Following KDA and Kaiser (2000) we can write:

\begin{equation}
E_{\rm tot}=u_{\rm c}V_{\rm c}=
\frac{2(5-\beta)}{9[\Gamma_{\rm c}+(\Gamma_{\rm c}-1)({\cal P}_{\rm hc}/4)]-4
-\beta}Q_{0}t.
\end{equation}

\noindent
Thus, the ratio of enery delivered by the twin jets and stored in the cocoon is:

\[\frac{2Q_{0}t}{E_{\rm tot}}=\frac{9\Gamma_{\rm c}-4-\beta}{5-\beta} +
\frac{9(\Gamma_{\rm c}-1)}{4(5-\beta)}{\cal P}_{\rm hc},\]

\noindent
i.e. if $\Gamma_{\rm c}$=const and $\beta$=const, 
this ratio is a function of the pressure ratio ${\cal P}_{\rm hc}$ only.
For $\Gamma_{\rm c}=5/3$ and $\beta=3/2$ we have

\begin{equation}
2Q_{0}t/E_{\rm tot}=2.7+0.43 {\cal P}_{\rm hc}.
\end{equation}

\subsection{The fitting procedure}

On the basis of the above model we aim to predict the specific physical
parameters of the sample {\sl giants} and normal-size sources at their estimated
(dynamical) ages. These are: $Q_{0}$, $\rho_{0}$, $u_{\rm c}$, $p_{\rm c}$, and
$E_{\rm tot}$. This differs from the KDA approach, who basing on
available observational data, evaluated some general trends and made crude
estimation of possible range of values attained by the model parameters.
Following them, we adopt their `Case~3' where both the cocoon and magnetic
field are `cold' ($\Gamma_{\rm c}=\Gamma_{\rm B}=5/3)$ and the adiabatic index
of the jet and internal gas is also $5/3$. For the initial ratio of the energy
densities of the magnetic field and the particles we use
$r\equiv u_{\rm B}/u_{\rm e}=(1+p)/4$, with the exponent of the energy
distribution $p=2.14$.

The core radius $a_{0}$ is one of the most difficult model parameter to be set
up. Even carefull 2-D modelling of a distribution of radio emission  for well
known sources with quite regular structures can lead to values of $a_{0}$
discrepant with those predicted by X-ray observations, the only presently
available method to determine the source environment (cf. an extensive discussion
of this problem in Kaiser 2000). In our statistical approach we assume
$a_{0}=10$ kpc for all sources, a conservative value in between 2 kpc used by
KDA and 50 kpc found by Wellman et al. (1997). In Sect.~7.1 we discuss
the consequences of other possible values of this parameter. We also use a
constant value of $\beta$ for all sample sources. This exponent can be easily
estimated from the slope of the observed relation between
log$\,B_{corr/z}$ and log$\,t$ in Fig.2b. As it is $-0.65\pm 0.04$,
then $a=(-0.65\times 2)\pm  0.08$ and $\beta=1.09\pm 0.20$. This slope is much
flatter than $\beta=1.9$ adopted in KDA on the basis of Canizares et al.'s
(1987) paper who found that value as typical for a galaxy to about 100 kpc
from its center. The latter value was probably justified because the
uncorrected for redshift relation log$\,B_{\rm eq}-$log$\,t$ has a steeper
slope of $-0.98\pm 0.05$ what implies $\beta=1.96\pm 0.21$. Thus, we take
$\beta=1.5$ for further calculations which is compatible
with other estimates of this parameter (e.g. Daly 1995).
Another free parameter of the model, the orientation of the jet axis in respect
to the observer's line-of-sight $\theta=90^{\circ}$ is assumed for all {\sl giants}
and $\theta=70^{\circ}$ for other sources. This latter value is justified by the 
dominance of FRII-type radio galaxies in our sample. In view of the unified 
scheme for extragalactic radio sources, an average orientation angle 
$\langle\theta_{\rm RG}\rangle\simeq 69^{\circ}$ for radio galaxies only was 
determined by Barthel (1989).

Given the values of $a_{0}$, $\beta$ and $\theta$, one can find the jet power
$Q_{0}$ and the initial density of external medium $\rho_{0}$ for each individual
sample source characterized by the observational parameters: the age $t$, axial
ratio $AR$ and redshift $z$, by fitting the jet length $L_{\rm j}$ and the
luminosity of cocoon to its observed values of $D/2$ and $P_{\nu}$, respectively.
The above fitting procedure was done iteratively and proved to give always
stable and unique solutions.

\section{Results of the modelling}

The fitted physical parameters for all the sample sources are given in the
Appendix. Below, we analyse a range of values and distributions of the above
parameters attained by the sample sources.

\subsection{Jet power $Q_{0}$ and core density $\rho_{0}$}

A distribution of these parameters on the log($Q_{0}$)--log($\rho_{0}$)
plane is shown in Fig.~1a. As both parameters should be independent
between themselves, we test whether the observed distribution is or is not
biased by possible selection effects. The data in Table~A2 implicate that
$Q_{0}$ correlates, in order of a significance level of the correlation, with
luminosity $P_{\rm 1.4}$, redshift $z$ and age $t$. Calculating the Pearson
partial correlation coefficients, we found no significant correlation
between $Q_{0}$ and $\rho_{0}$ when $z$ (or $P_{\rm 1.4}$) and $t$ are kept
constant.

\begin{figure}
\resizebox{\hsize}{!}{\includegraphics{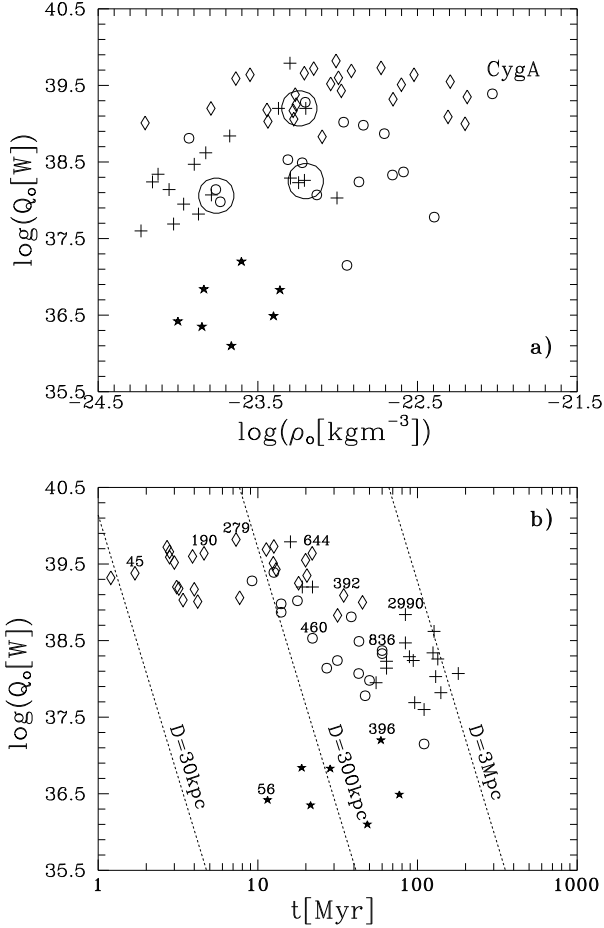}}
\caption[]{Plots of the jet power $Q_{0}$ {\bf a)} against central density of
the core $\rho_{0}$; {\bf b)} against source age. The {\sl giants} are indicated
by crosses, high-redshift sources -- diamonds, low-redshift sources -- open
circles, and low-redshift/low-luminosity sources -- stars. The `clans' of sources 
with
similar $Q_{0}$ and $\rho_{0}$ are marked in {\bf a)} by the large circles. The
dotted lines in {\bf b)} mark a constant linear size predicted from Eq.(6). The
numbers indicate actual size of the source beneath}
\end{figure}

In view of the above, one can realise in Fig.~1a that (i) among the sources with
a comparable jet power $Q_{0}$, {\sl giant} sources have an average central
density $\rho_{0}$ smaller than a corresponding central density of normal-size
sources, (ii) {\sl giants} have at least ten times more powerful jets than much
smaller low-luminosity sources of a comparable $\rho_{0}$, and (iii) for
a number of sources in the sample the derived values of their fundamental
parameters $Q_{0}$ and $\rho_{0}$ are very close, while their ages are
significantly different. Thus in view of the model assumptions, they may be
 considered as `the same' source observed at different epochs of its
lifetime. Such bunches of three to five sources (hereafter called `clans') are
indicated in Fig.~1a with the large circles. These clans have appeared crucial
in a comparison of the observing data and the model predictions, and in the
analysis of the {\sl giant}-source phenomenon. More detailed analysis of these
clans and their evolution is given in Sect.~6.2.

\subsection{Relations between principal parameters of the sources}
 
In this subsection we analyse a degree of correlation between the principal
derived parameters of the sources: $Q_{0}$, $\rho_{0}$, $B_{\rm eq}$, and their 
linear size $D$, axial ratio $AR$, redshift $z$, and
assumed age $t$. For the reason that most relations between different parameters
seem to be a power law,  hereafter we analyse their linear correlations in the
logarithmic form. Slopes of these correlations are then used to constrain the
dynamical evolution model outlined in the previous Section.
 
\subsubsection{The relations between log($D$), log($Q_{0}$), log($\rho_{0}$), 
log($t$), and log($1+z$)}
 
A strong correlation between linear size and spectral age of 3C radio sources was
already noted by Alexander \& Leahy (1987) and confirmed by Liu et al. (1992).
This correlation in our samples is shown in Fig.~2a. The {\sl giant} sources
do not show any departure from the correlation for the comparison sources.
The same conclusion has been made by Schoenmakers et al. (2000). This means that
the mean expansion velocities of those sources are similar. However, two other
details are worth to emphasize: (i) There are four high-redshift {\sl giants}
which are much younger sources than the low-redshift {\sl giants}. Two of them
are quasars. It seems that they might grow so large under some exceptional
conditions. (ii) The (log) $D-t$ relation for the low-luminosity (mostly B2)
sources follows a slope of the correlation, but those sources are definitely much
smaller indicating a dependence of the size and expansion velocity on luminosity.
 
We also note a very significant anticorrelation between redshift and age. This
anticorrelation is expected because older sources at higher redshifts may fall
down beyond a survey flux-limit due to the spectral and, especially inverse
Compton, losses and also adiabatic expansion (cf. Blundell et al.). Since
each parameter of sources in our sample correlates somehow with other parameters,
in order to determine which correlation is the strongest, we calculate the
Pearson partial correlation coefficients between selected parameters.
All correlations are calculated between given
parameters taken in the logarithmic scale (for the sake of simplicity, the `log'
signs are omitted in the Tables below, showing these correlations). Hereafter
$r_{XY}$ is the correlation coefficient between parameters
$X$ and $Y$, $r_{XY/W}$ is the partial correlation coefficient between these
parameters in the presence of a third one, ($W$), which can correlate with both
$X$ and $Y$, and $P_{XY/W}$ is the probability that the test pair $X$ and $Y$ is
uncorrelated when $W$ is held constant. Similarly, $r_{XY/VW}$, $r_{XY/UVW}$,
$P_{XY/VW}$, and $P_{XY/UVW}$ are the correlation coefficients for the correlations
involving four or five parameters, and the related probabilities, respectively.

\begin{figure}
\resizebox{\hsize}{!}{\includegraphics{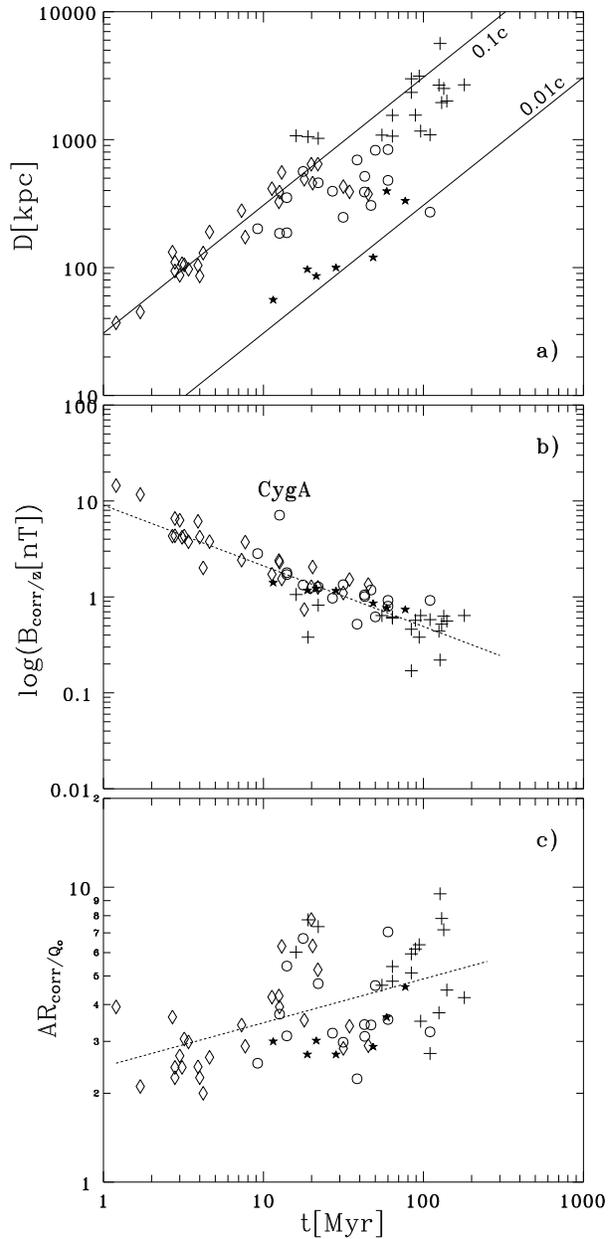}}
\caption[]{A distribution of selected observational parameters against the age:
{\bf a)} projected linear size $D$; {\bf b)} redshift-corrected equipartition
magnetic-field strength $B_{corr/z}$ (cf. the text); and {\bf c)} jet 
power-corrected axial ratio $AR$. The symbols indicating sources within different 
groups are the same as in Fig.~1. The solid lines in {\bf a)} indicate an 
expansion velocity of $0.1c$ and $0.01c$. The dashed lines in {\bf b} and {\bf c} 
indicate the linear regressions on the age axis}
\end{figure}

The partial correlation coefficients between selected combinations of the
source parameters with the related probabilities of their chance correlation
are given in Table~1.
 
\begin{table*}[htb]
\caption{The correlations between (log) $D$ and $t$ or $Q_{0}$ or $\rho_{0}$
when other parameters are kept constant}
\begin{tabular}{@{}lllll}
\hline
Correlation & $r_{XY/U}$  & $P_{XY/U}$\\
            & $r_{XY/V}$  & $P_{XY/V}$\\
            & $r_{XY/W}$  & $P_{XY/W}$  & $r_{XY/UVW}$  & $P_{XY/UVW}$\\
\hline
$D-t$/$Q_{0}$        & +0.943  & $\ll$0.001\\
$D-t$/$\rho_{0}$     & +0.807  & $\ll$0.001\\
$D-t$/1+$z$          & +0.840  & $\ll$0.001\\
$D-t$/$Q_{0}$,$\rho_{0}$,1+$z$  & &     & +0.959  & $\ll$0.001\\
& &\\
$D-Q_{0}$/$\rho_{0}$ & +0.013  & 0.918\\
$D-Q_{0}$/$t$        & +0.820  & $\ll$0.001\\
$D-Q_{0}$/1+$z$      & +0.459  & $\ll$0.001\\
$D-Q_{0}$/$\rho_{0}$,$t$,1+$z$ & &      & +0.884  & $\ll$0.001\\
& &\\
$D-\rho_{0}$/$Q_{0}$ & $-$0.194 & 0.119\\
$D-\rho_{0}$/$t$     & $-$0.080 & 0.525\\
$D-\rho_{0}$/1+$z$   & $-$0.086 & 0.492\\
$D-\rho_{0}$/$Q_{0}$,$t$,1+$z$  & &     & $-$0.745 & $\ll$0.001\\
\hline
\end{tabular}
\end{table*}
 
\noindent
In view of the dynamical model applied and as a result of the above statistical
correlations, we realize that the linear size of a source strongly depends on
both its age and the jet power, where the correlation with age is the strongest.
However, the size also anti-correlates with central density of the core. That
anticorrelation seemes to be a weaker than the correlations with $Q_{0}$ and $t$  
and become well pronounced only when
all three remaining parameters ($Q_{0}$, $t$ and $z$) are kept constant.

All dynamical models predict that luminosities of matured radio sources decrease
with their age. Therefore, more distant sources fall through the flux-density
limit of a sample sooner than nearer sources, and in any sample the 
high-redshift sources will be younger and more luminous than the
low-redshift ones.
A  significant anticorrelation between $Q_{0}$ and age $t$, expected as a 
consequence of the above effect (called `youth-redshift degeneracy' by Blundell 
et al.), is shown in Fig.~1b.

\subsubsection{The relation between log($B_{\rm eq}$), log($t$), and
log(1+$z$)}
 
An anticorrelation between $B_{\rm eq}$ and $t$ is expected due to the basic
relations for energy densities defined in Sect.~4.2.
The Pearson partial correlation coefficients and the relevant probability for
the chance correlation between $B_{\rm eq}$ and $t$,
independent of redshift, are given in Table~2.
 
\begin{table}[thb]
\caption{The correlation between (log) $B_{\rm eq}$, $t$, and 1+$z$}
\begin{tabular}{@{}lccr}
\hline
Correlation     &$r_{XY}$   &$r_{XY/W}$ & $P_{XY/W}$\\
\hline
$B_{\rm eq}-t$   & $-$0.921  & $-$0.815  & $\ll$0.001\\
$B_{\rm eq}-$(1+z) & +0.750  & +0.191    & 0.125\\
(1+z)$-t$       & $-$0.762  & $-$0.278  & 0.024\\
\hline
\end{tabular}
\end{table}
 
Table~2 confirms that there is a strong and very significant anticorrelation
between the equipartition magnetic field and age of the lobes or cocoon. An
apparent correlation between $B_{\rm eq}$ and 1+z is less significant. A rate
of decrement of the magnetic field strength is an indication of the dynamical
evolution of a source (cf. Sect.~4).  Thus, in order to determine a slope
 of the linear dependence between log($B_{\rm eq}$) and log($t$)
independent of redshift, we have transformed the $B_{\rm eq}$ values from
Table~A.2 to a reference redshift of 0.5 using the correlation between
log($B_{\rm eq}$) and log(1+z). The relation between the transformed magnetic 
field in the sources and their age is shown in Fig.~2b.
 
\subsubsection{The relation between log($AR$), log($Q_{0}$), and log($t$)}
 
A first time (to our knowledge) axial ratios of {\sl giant} sources were
analysed and compared with those of smaller FRII-type sources by Subrahmanyan et
al. (1996), who found no difference between the axial ratios of eight {\sl
giants} and eight 3C sources with a median size of about 400 kpc. The authors
did not precise what 3C sources were considered, but since all {\sl giant} and
normal sources were of comparable powers and at comparable redshifts, we
pressume they might be at similar ages, so the dependence of $AR$ on time could
not be visible.
 
In the model of Blundell et al., the axial ratio of an individual source
steadily increases throughout its lifetime. Besides, that model implicates a
dependence of the $AR$ on the jet power $Q_{0}$. The latter dependence was
probably reflected by an apparent correlation between $AR$ and 178-MHz luminosity
of 3C sources noted by Leahy \& Williams (1984). Taking into account the
unavoidable anticorrelation between $Q_{0}$ and redshift (cf. Sect.~5.2.1), in
Table~3 we
have calculated the partial correlation coefficients and the related
probabilities of chance correlations between $AR$ and $t$, $AR$ and $Q_{0}$, and
$AR$ and $\rho_{0}$ when relevant combinations of the parameters $t$, $Q_{0}$,
$\rho_{0}$, and 1+$z$ are kept constant. 
  
\begin{table*}[htb]
\caption{The correlations between (log) $AR$ and $t$ or $Q_{0}$ or $\rho_{0}$
when other parameters are kept constant}
\begin{tabular}{@{}lllll}
\hline
Correlation & $r_{XY/U}$  & $P_{XY/U}$\\
            & $r_{XY/V}$  & $P_{XY/V}$\\
            & $r_{XY/W}$  & $P_{XY/W}$  & $r_{XY/UVW}$  & $P_{XY/UVW}$\\
\hline
$AR-t$/$Q_{0}$    & +0.625  & $\ll$0.001\\
$AR-t$/$\rho_{0}$ & +0.456  & $\ll$0.001\\
$AR-t$/1+$z$      & +0.493  & $\ll$0.001\\
$AR-t$/$Q_{0}$,$\rho_{0}$,1+$z$ & &      & +0.517  & $\ll$0.001\\
& &\\
$AR-Q_{0}$/$t$      & +0.562 & $\ll$0.001\\
$AR-Q_{0}$/$\rho_{0}$ & +0.116 & 0.353\\
$AR-Q_{0}$/1+$z$   & +0.447 & $\ll$0.001\\
$AR-Q_{0}$/$t$,$\rho_{0}$,1+$z$  & &     & +0.424 & $\ll$0.001\\
& &\\
$AR-\rho_{0}$/$t$        & +0.357  & 0.003\\
$AR-\rho_{0}$/$Q_{0}$    & +0.180  & 0.148\\
$AR-\rho_{0}$/1+$z$      & +0.290  & 0.018\\
$AR-\rho_{0}$/$t$,$Q_{0}$,1+$z$  & &     & +0.186  & 0.142\\
\hline
\end{tabular}
\end{table*}

Table~3 shows a statistically significant correlations between the axial ratio
and the source's age, as well as the axial ratio and the jet power. Therefore, we
have fitted a plane to the values of log($AR$) over the (log)  $Q_{0}$--$t$
plane. In this and other fits of a plane performed hereafter, the ordinate values
are weighted by the square of their uncertainty given in Tables A1 or A2.
As a result of the above fit we found
 
\begin{equation} 
AR(Q_{0},t)\propto Q_{0}^{0.12\pm 0.02}t^{0.23\pm 0.04}.
\end{equation}

\noindent
Indeed, the statistical data strongly support the implication of the Blundell
et al.'s model about a dependence of $AR$ on $Q_{0}$. A consequence of this
effect for the expansion speed of the cocoon is pointed out in the next
subsection.
Using the above relation, one can transform the apparent $AR$ values from 
Table~A.1 to a reference jet power of $10^{39}$\,W. The relation between the
transformed axial ratio and age of the sample sources is shown in Fig.~2c.

\subsection{Statistical expansion speed of the cocoon}

The expansion speed of the cocoon along the jet axis is indirectly described by
Eq.(2). Inserting $D=2L_{\rm j}$ and $AR=2R_{\rm T}$ into Eq.(2), we have
$D(t)\propto (AR)^{2/3}t^{3/(5-\beta)}$. Substitution of $AR$ from Eq.(5) and
$\beta=1.5$ into Eq.(2) gives $D$ as a function of $Q_{0}$ and $t$, i.e. the
expansion speed dependent on the jet power $Q_{0}$

\begin{equation}
D(Q_{0},t)=c_{1}\,Q_{0}^{0.10\pm 0.03}t^{1.08\pm 0.04},
\end{equation}

\noindent
where $c_{1}$ is a constant to be empirically determined from the observed data
of size and age. Expressing $D$ in kpc and $t$ in Myr, the constant
$c_{1}=(2.2\pm 0.04)\,10^{-3}$ has been found.
The power exponent over $t$ greater than unity implicates that the cocoon
may expand in time with some systematic though minor acceleration. Moreover,
this acceleration [according to Eq.(6)] depends on the jet power. For example,
if $Q_{0}=10^{39}$ W, an average time needed to expand the cocoon from
30 to 300 kpc is about 12.1 Myr which gives an expansion speed of about
0.073$c$, while about 28.2 Myr is necessary to expand it from 300 kpc to 1 Mpc
which gives the speed of 0.081$c$. If the jet power is two orders fainter
($Q_{0}=10^{37}$ W), these speeds are reduced to 0.047 and 0.053$c$,
respectively.
The relation between $AR$ and $t$ in Eq.(5) implicates a time {\sl deceleration}
of the cocoon's expansion if a decrease of the environment density with distance
from the cluster centre is slower, i.e. if $\beta<1.1$. However, a time 
evolution of the, so called, `clan sources' (Sect.~6.2) seems to support the
possibility of a slow acceleration in their case.

 A distribution of the sample sources on the $\log Q_{0}-\log t$
plane (Fig.~1b) shows a dramatic discrepancy between the observed large size and
derived short lifetime of three {\sl giants}: 3C274.1, 3C292, and 0437$-$244.
They are three of four high-redshift {\sl giant}
sources in our sample. If their ages are real, they must expand with the highest
velocity of about $0.2\sim 0.25c$ (cf. also Fig.~2a). More extensive discussion
of the above effect is given in Sect.~7.3.

\begin{figure}
\resizebox{\hsize}{!}{\includegraphics{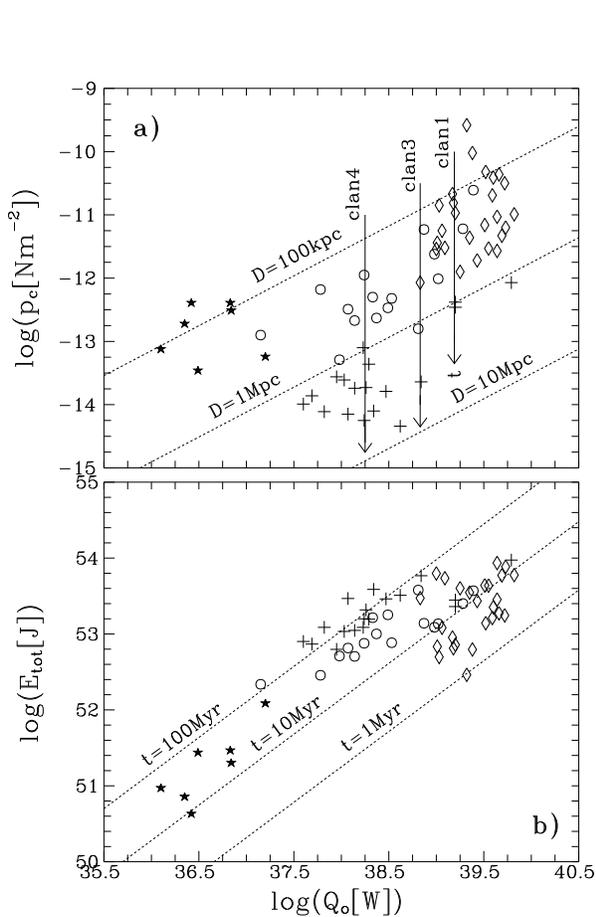}}
\caption{A distribution of modelled parameters against the jet power: {\bf a)}
the cocoon pressure $p_{\rm c}$; {\bf b)} the total energy of the cocoon
$E_{\rm tot}$.
The symbols indicating sources within different groups are the same as in Figs.1
and 2. The dotted lines mark in {\bf a)} a constant linear size of source from
the correlation $\log p_{\rm c}=(-1.79\pm 0.04)\log D[{\rm kpc}]+(0.76\pm 0.02)
\log Q_{0}[{\rm W}]-(36.9\pm 0.9)$, in {\bf b)} -- a constant age from the correlation
$\log E_{\rm tot}=(0.90\pm 0.03)\log t[{\rm Myr}]+(0.94\pm 0.02)\log Q_{0}[{\rm W}]+
(15.6\pm 0.5)$.
Three vertical lines in {\bf a)} indicate time axes passing
through members of the `clans' of sources with very close values of $Q_{0}$
and $\rho_{0}$ (cf. the text)}
\end{figure}

\subsection{Pressure and energy of the cocoon}

In the model, energetics of the radio source is governed by the jet power $Q_{0}$,
adiabatic index $\Gamma_{\rm c}$, and the pressure ratio ${\cal P}_{\rm hc}$. Since
$\Gamma_{\rm c}$ is assumed constant for all the sources and $Q_{0}$ is constant
for a given source, the energy of the cocoon $u_{\rm c}$ and its total energy
$E_{\rm tot}$ are determined by $p_{\rm c}$ attained by the source at age $t$.
The cocoon pressure of the sample sources is plotted against $Q_{0}$ in Fig.~3a.
It is clearly seen that the {\sl giants} have the lowest cocoon pressure and,
hence, the lowest energy density among all sources in the sample. As $p_{\rm c}$
decreases with $t$, and the size of source (cocoon) $D$ increases with $t$, we
have calculated the partial correlation coefficients for the correlation
between log($p_{\rm c}$) and log($Q_{0}$) at constant $t$ or constant $D$. In a
result, this correlation in our sample is much stronger for $D$=const
($r_{XY/W}=+0.969$) than that for $t$=const ($r_{XY/W}=+0.379$). Therefore, the
lines of constant $D$ in Fig.3a almost perfectly agree with actual sizes of the
sources. The vertical lines indicate the time axis for members of the clans
introduced in Sect.~5.1 and whose dynamical evolution is analysed in Sect.~6.2
and 6.3.

\begin{figure}
\resizebox{\hsize}{!}{\includegraphics{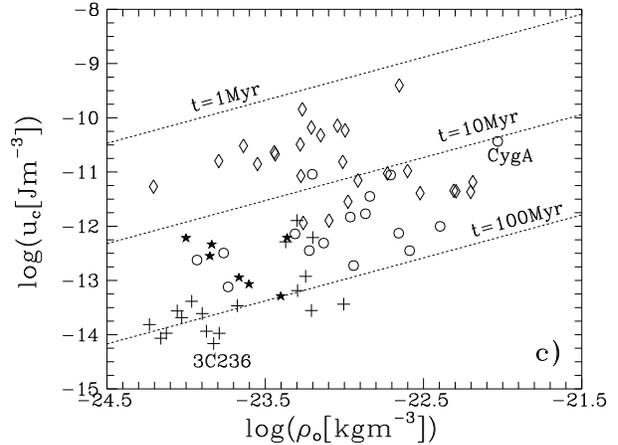}}
\caption[]{Energy density of the cocoon $u_{\rm c}$ against the central core
density $\rho_{0}$. The symbols -- as in Fig.3. The dotted lines indicate a
constant age from the statistical correlation in the sample:
$\log u_{\rm c}=(-1.87\pm 0.04)\log t[{\rm Myr}]+(0.80\pm 0.04)\log\rho_{0}+
(9.3\pm 0.9)$}
\end{figure}

In Fig.~3b, a total energy of the cocoon $E_{\rm tot}$ is plotted against $Q_{0}$.
It shows that the energy contained in {\sl giant} sources spans the same range
of values as does energy of normal-size sources in spite of their redshift,
luminosity or age. This implicates that the decrease of energy density in time
is compensated by a parallel increase of the source (cocoon) volume. Only the
low-luminosity sources have energies much smaller. Imagining a division of the
log($E_{\rm tot}$)--log($Q_{0}$) plane into some quadrants, the {\sl giant},
and normal-size low-redshift and high-redshift sources are in a quadrant with
$Q_{0}>10^{37.5}$ W and $E_{\rm tot}>10^{52.5}$ J, while the low-luminosity
low-redshift radio galaxies are in the quite opposite quadrant with
$Q_{0}<10^{37.5}$ W and $E_{\rm tot}<10^{52.5}$ J. Since the relation between
$E_{\rm tot}$ and $Q_{0}$ at constant $t$, dependent only of ${\cal P}_{\rm hc}$,
is very high (the corresponding partial correlation coefficient equals to
+0.991), the dotted lines in Fig.~3b indicating a
constant age of sources also agree very well with the data. All {\sl giants}
but four high-redshift ones have the highest lifetime.

Now, analysing how the energetics of the sources relates to the core density
$\rho_{0}$, we found that a pressure in the cocoon $p_{\rm c}$ (thus the energy
density $u_{\rm c}$ as well) also significantly depends on $\rho_{0}$. A
distribution of the sample sources on the log($u_{\rm c}$)--log($\rho_{0}$)
plane is shown in Fig.~4. Although the correlation coefficient between
log($u_{\rm c}$) and log($\rho_{0}$) in Fig.~4 is only +0.548, a calculation
shows that this direct correlation is smoothed by other correlations between
the parameters $u_{\rm c}$, $\rho_{0}$, $t$, and 1+$z$. The partial correlation
coefficient for the correlation between log($u_{\rm c}$) and log($\rho_{0}$),
when $t$ and 1+$z$ are constant, equals to +0.932. Therefore the dotted lines of
constant age in Fig.~4 again corresponds very well to the actual ages of the
sources. On this plane giant sources occupy an area of the lowest values of
$u_{\rm c}$ and $\rho_{0}$.

\subsection{Energy ratio $2Q_{0}t/U_{\rm eq}$}

The ratio of total energy supplied by twin jets during the lifetime of a
source $(2Q_{0}t)$ and its energy stored in the cocoon, derived from the data
under the equipartition assumption, $U_{\rm eq}=u_{\rm eq}V_{\rm c}$, allows
another test of the dynamical model predictions, and is given in column 9 of
Table~A2. This ratio vs. the cocoon axial ratio $AR$ is plotted in Fig.~5. The
uncertainties of both values are marked by error bars on some data points.
The solid curve indicates the model prediction from Eq.(4), while the dashed
curve shows the best fit to the weighted data. The observed trend fully 
corresponds to
the model prediction. However, the derived values of $2Q_{0}t/U_{\rm eq}$, i.e.
the reciprocal of an efficiency factor by which the kinematic energy of the jets
is converted into radiation, is much higher than $\approx$2, a value usually
assumed in a number of papers. This aspect is discussed in Sect.~7.4.

\begin{figure}
\resizebox{\hsize}{!}{\includegraphics{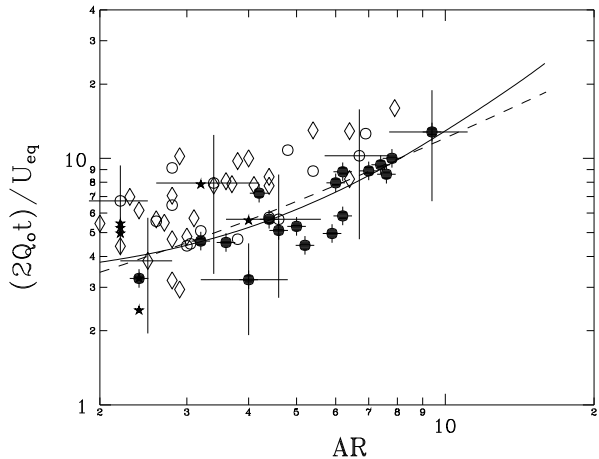}}
\caption[]{Ratio of the total energy supplied by twin jets and energy stored
in the cocoon against its axial ratio $AR$. The large uncertainties of both
parameters are marked by error bars for a few sources. Here the giant sources
are marked with filled circles. The solid curve
indicates the model prediction from Eq.(4); the dashed curve shows the best
fit to weighted data}
\end{figure}

\subsection{The correlation of the cocoon pressure with age and redshift}

In our statistical analysis we have noted a significant correlation between the
cocoon pressure and age of sources. But age also correlates with redshift, so we
have calculated the partial correlation between these parameters. Since dynamical
age of sources in our sample seems to be reliably estimated, we can show that
the cocoon pressure $p_{\rm c}$ very weakly depends on redshift but very strongly
upon the source age. The partial correlation coefficients between $p_{\rm c}$,
$t$, and 1+$z$ (all values in logarithmic scale) are given in Table 4.

\begin{table}[htb]
\caption{The correlation between (log) $p_{\rm c}$, $t$, and 1+$z$}
\begin{tabular}{@{}lccr}
\hline
Correlation       & $r_{XY}$  & $r_{XY/W}$  & $P_{XY/W}$\\
\hline
$p_{\rm c}-t$     & $-$0.924    & $-$0.817  & $\ll$0.001\\
$p_{\rm c}-$(1+z) & +0.771      & +0.267    & 0.300\\
(1+z)$-t$         & $-$0.762    & $-$0.206  & 0.097\\
\hline
\end{tabular}
\end{table}

Thus, Table~4 shows that the apparent correlation between the cocoon pressure
and redshift is caused by the stronger correlations between the pressure and
age, and between the age and redshift. Fitting a plane to the values of log
$p_{\rm c}$ over the (log) $t$ -- (1+$z$) plane  we found

\[\log p_{\rm c}=(-1.79\pm 0.15)\log t[{\rm Myr}]+(1.56\pm 0.65)\log(1+z)-\]
\[(10.0\pm 0.30).\]

Therefore, one can transform the pressure values either to a fixed age or to a
constant redshift. The plot of cocoon pressures transformed to $z=0.5$
[$p_{\rm corr/z}$] versus age $t$ is shown in Fig.~6a, while the pressures
transformed to $t=10$ Myr versus (1+$z$) [$p_{\rm corr/t}$] -- in Fig.~6b.
These plots well
illustrate very different partial correlation coefficients $r_{XY/W}$ given in
Table 4. The above result put a new insight into a cosmological evolution of the
IGM and is discussed in Sect.~7.5.

\begin{figure*}
\resizebox{\hsize}{!}{\includegraphics{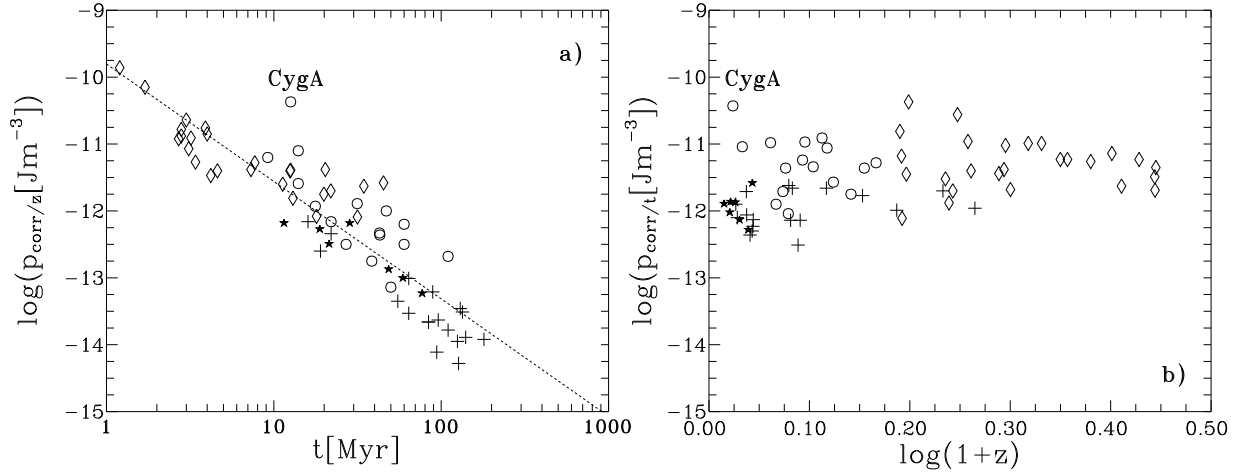}}
\caption[]{{\bf a)} Cocoon pressure transformed to the constant redshift of 0.5
against age of the sample sources. The best linear fit is indicated by the
dotted line; {\bf b)} The same pressure transformed to the constant age of 10
Myr against redshift}
\end{figure*}

\section{Evolutionary tracks of sources}

The KDA model  enables one to calculate evolutionary tracks of some
secondary parameters of radio sources if their primary parameters $Q_{0}$ and
$\rho_{0}$ are given. In the papers of KDA and Blundell et al.
 the tracks of radio luminosity $P$ versus linear size $D$ were derived
for imaginary sources with assumed values of $Q_{0}$, $\rho_{0}$, $a_{0}$,
$\beta$, and $z$.

In our approach we are able to calculate such evolutionary tracks for actual
sources. In Sect.~5.1, the `clans' of sources have been pointed out,
 i.e. the sources with
very close values of both fundamental parameters: $Q_{0}$ and $\rho_{0}$, and
evidently different ages and axial ratios. Since the dynamical model assumes
constant jet power during a source lifetime, and a nucleus density $\rho_{0}$
is {\sl a priori} constant, members of such clan can be considered as `the same'
source observed at a number of different epochs throughout its life.
The observed parameters of these members can verify predictions of the model.
However, fits of the tracks predicted with the original KDA model to the observed
parameters of sources have appeared unsatisfactory. Much better fits of the
modelled tracks to observational data of the `clan' members are found with the
cocoon axial ratio evolving in time, as reported in Sect.5.3. This, in turn,
implicates a time evolution of the pressure ratio ${\cal P}_{\rm hc}$. Indeed,
substitution of Eq.(5) into Eq.(1), one has (for $\beta=3/2$, and where $Q_{0}$ is
in watts and $t$ in Myr)

\begin{equation}
{\cal P}_{\rm hc}(t)\approx 2.3\,10^{-8}Q_{0}^{0.20\pm 0.03}t^{0.38\pm 0.05}.
\end{equation}

\subsection{The clans}

We found 6 clans consisting of 3, 4 or 5 sample sources fulfilling a selection
criterion, namely that the fitted values of $Q_{0}$ and $\rho_{0}$ do not differ
between themselves by more than 30 per cent. Also redshift of members should be
comparable, however in our small sample we have accepted the redshift ratios up
to about 3. Three of six clans are marked in Fig.~3a. Their members are
listed in Table 5.

\begin{table}[htb]
\caption{Members of the selected clans}
\begin{tabular}{@{}llll}
\hline
               & Clan1   & Clan3   & Clan4\\
\hline
$\langle\lg Q_{0}\rangle$ & 39.26 & 38.84 & 38.21\\
$\langle\lg\rho_{0}\rangle$ & $-$23.24 & $-$23.35 & $-$23.22\\
redshift range & 1:2.3   & 1:2.3   & 1:3.4\\
members        & 3C263.1 & 3C166   & 3C319\\
               & 3C289   & 3C334   & 1025$-$224\\
               & 3C411   & 1012+488& 0136+396\\
               & 3C272   & 0821+695& 3C326\\
               & 3C274.1\\
\hline
\end{tabular}
\end{table}

\begin{figure}
\resizebox{\hsize}{!}{\includegraphics{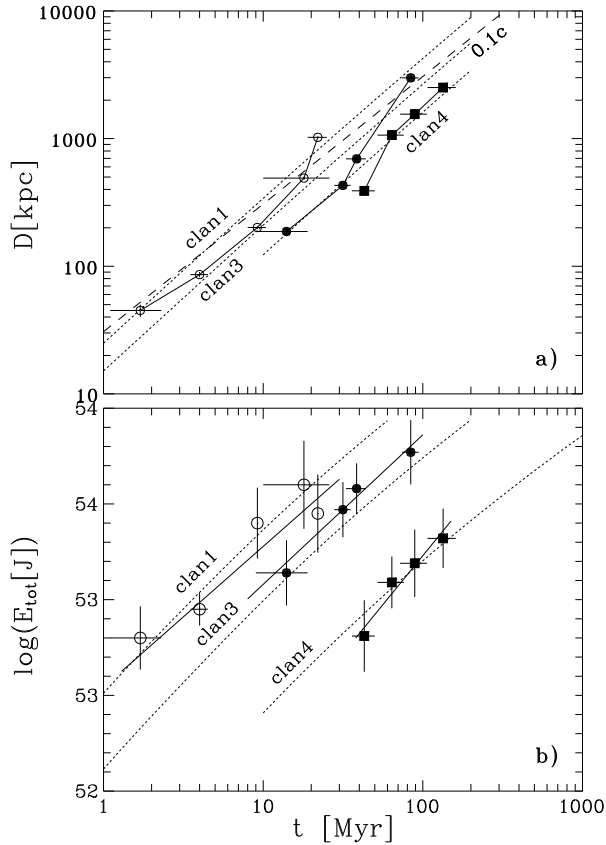}}
\caption[]{{\bf a)} Time evolution of the size of a fiducial source for each of
the three observed clans (dotted lines). The dashed line indicates a constant
advance speed of 0.1$c$; {\bf b)} Evolution of the total energy of a fiducial
source for each of the clans. The shorter solid lines show the best linear fit
to the data points}
\end{figure}

\noindent
The members are ordered according to their increasing age. In all three clans
that increase of age is accompanied with an increase of the size and total
energy of the members, altogether with a decrease of their luminosity and energy
density. This qualitative observational behaviour, concordant with the model
expectations, is quantitatively presented in the following subsections.

\subsection{Time evolution of the clans}

In order to check whether observed parameters of the members of a given clan
are consistent with a time evolution derived from the model, a fiducial source
has been determined for each clan. This fiducial
source has $Q_{0}$ and $\rho_{0}$ equal to the mean value of these parameters
in a given clan and the evolving axial ratio of the cocoon, as given by Eq.(5).
For each of the clans, a time evolution of its size $D(t)$, luminosity $P(t)$,
energy density $u_{\rm c}(t)$, and total energy $E_{\rm tot}(t)$
have been calculated. The predicted $D(t)$ tracks for the three clans are shown
in Fig.~7a with the dotted lines overlayed on the solid lines connecting members
of each clan drawn with different symbols. The data points are shown with error
bars (the errors in size are usually smaller than the vertical size of symbols).
The dashed line indicates $D(t)$ for the constant advance speed of 0.1$c$. 
The higher slopes of the modelled tracks in respect to this constant speed seem
to confirm accelerating a little expansion of the cocoon in the investigated
clans (as deduced in Sect.~5.4), however the effect is of low significance 
because of uncertainties of the age. 

The predicted evolution of the total energy for the three clans is shown in
Fig.~7b with the dotted lines. The solid lines indicate the best linear fit to
the data points. A correspondence between the modelled $E_{\rm tot}(t)$
and the data points is satisfactory.

\subsection{The tracks log$P$--log$D$}

The evolutionary tracks for  the three clans are shown in Fig.~8 with solid
curves. The markers of the same time on these tracks are connected with a dotted
line. The members of separate clans are indicated by different symbols. A numer
in vicinity of each symbol indicates actual age of the corresponding source in
Myr. The dashed curves show the relevant track calculated from the original KDA,
model, i.e. with a constant $AR$ taken as the mean of axial ratios in a given
`clan'. It is clearly seen that the evolving $AR$
much better fits the observed changes of $P$ and $D$. The further discussion
about these tracks is given in Sect.~7.6.

\subsection{The tracks log($u_{\rm eq}$)--log($E_{\rm tot}$)}

The model also allows to predict the evolution of a source on the energy
density--total energy plane. These tracks are shown in Fig.~9 with solid
curves. As in Fig.~8, the time markers are connected with dotted lines, and
the member sources of each clan are indicated with the same symbols. Note that
the ordinate of every source is its equipartition energy density derived from
the observational data, while the modelled tracks show the {\sl fitted} cocoon
energy density. This is an additional argument that the modified model
satisfactory reproduce the data. In Fig.~9, an uncertainty of a value of both
parameters is shown by the error bars, where error of $u_{\rm eq}$ is given in
Table~A2 and error of $E_{\rm tot}$ is estimated from the error of source
volume. Again,
the numbers mark actual age of the corresponding source. The straight dashed
lines indicate the tracks computed with the constant $AR$. It is worth to
emphasize that the predicted evolutionary $u_{\rm c}$--$E_{\rm tot}$ tracks are
steeper and curved in respect to those expected from the original KDA
model. This effect, related to a rate of adiabatic losses and inflation of the
cocoon, is discussed in Sect.~7.6.

\begin{figure}
\resizebox{\hsize}{!}{\includegraphics{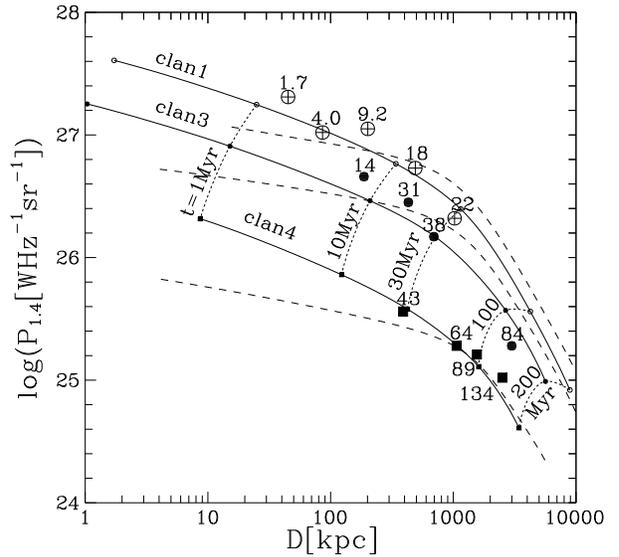}}
\caption[]{Evolutionary $P$--$D$ tracks fitted for three clans of sources with
evolving axial ratio $AR$ (solid curves). The markers of the same predicted age 
on each curve are connected with dotted lines. The members of each clan are 
marked with different symbols as those in Figs.~7a and 7b. Their actual
age is indicated by a number behind the symbol. The dashed curves indicate
relevant tracks but calculate with a constant $AR$, as in original KDA model}
\end{figure}

\begin{figure}
\resizebox{\hsize}{!}{\includegraphics{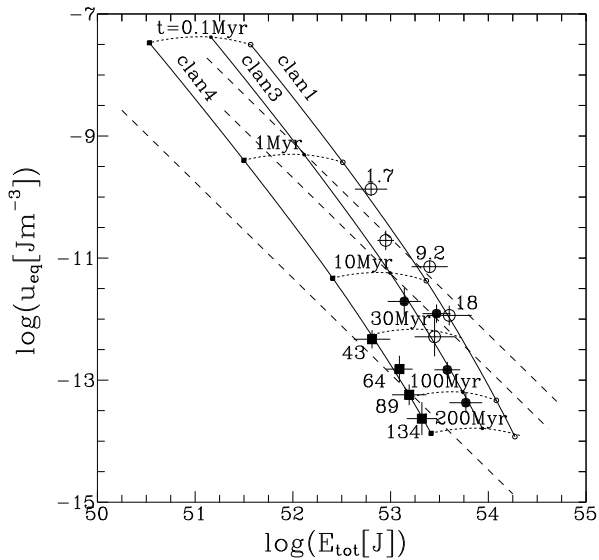}}
\caption[]{Tracks of evolving energy fitted for the three clans. All the lines
and symbols are the same as in Fig.~8.}
\end{figure}

\section{Discussion of the results and conclusions}

\subsection{Influence of fixed model parameters on the model predictions}

Basic physical parameters of the radio sources derived in this paper, i.e. jet
power $Q_{0}$, central density $\rho_{0}$, and cocoon pressure $p_{\rm c}$, are
in principle dependent on assumed central core radius, exponent in the external
gas distribution, adiabatic indices of electrons and magnetic field, as well as
on orientation of the jet axis towards the observer. In application of the KA
model we have assumed the same values of these parameters for all sample
sources. This assumption can only be valid in our statistical analysis of the
evolutionary trends in the whole FRII-type population but not for individual
sample sources. In particular, we have adopted $a_{0}=10$ kpc and $\beta=1.5$,
respectively.   Taking up a lower
core radius, e.g. $a_{0}=2$ kpc, and keeping $\beta=1.5$ will result in increase
of the modelled density of the core $\rho_{0}$ by roughly one order of
magnitude, while other modelled parameters will not be changed. Oppositely, a
lower density gradient, e.g. $\beta=1.1$, will lower $\rho_{0}$ approximately
$1.5\sim 6$ times. In this case however, the jet power will be increased about a
few percent and accordingly will be changed the pressure and energy density in
the cocoon. A relativistic equation of state for the magnetic field does not
change our results significantly, unless the KDA `Case~1' ($\Gamma_{\rm c}=\Gamma_
{\rm B}=4/3$) is adopted. However, this case, i.e. when both, the cocoon and the
magnetic field energy, have a relativistic equation of state, is unlikely for
our sample sources. Therefore, we argue that the results discussed below are not
significantly biased by a selection of particular set of the model parameters.

\subsection{A cause of extremal linear size}

In view of the KA and KDA analytical models of dynamical evolution of FRII-type
radio sources, many such sources can evolve into a stage characterized by the
linear size exceeding 1 Mpc. An access to this stage depends on a number of the
model parameters: jet power $Q_{0}$, its Lorentz factor $\gamma_{\rm jet}$,
adiabatic indices of the cocoon material and magnetic field, $\Gamma_{\rm c}$
nad $\Gamma_{\rm B}$, respectively, as well as a core radius $a_{0}$, external
gas density $\rho_{0}$, and exponent of its distribution $\beta$. For a given
set of these parameters, the model allows to determine whether an evolving
source will reach or not the size of 1 Mpc, and if so, at what age. From this
point of view, {\sl giant} sources should be the oldest ones.

In this paper we have confronted the model predictions with the observational
data on {\sl giant}-size and normal-size FRII sources. Our analysis strongly
suggests
that there is not a single evolutionary scheme governing the size development.
An old age or a low external density alone is insufficient to assure extremely
large linear extent of a source, both are necessary together with a suitable
power driven from AGN by the highly relativistic jets. About 83 per cent of 
{\sl giants}
in our sample possess the projected linear size over 1 Mpc owing to statistically
old age, low or moderate density of the external medium, and high enough power
of their jets. The remaining 17 per cent (3 sources) are high luminosity radio
galaxies at redshifts $z>0.4\sim 0.5$ and ages from 15 to 25 Myr which are
typical for normal-size sources. The jet power of these galaxies
is high enough to compensate for a higher ram pressure in a denser
surrounding environment and higher energy losses during the cocoon expansion.
The jet power of {\sl giants} is not extreme, so several FRII-type sources
having that $Q_{0}$ can potentially achieve very large size after a suitable
long time.
According to our results (cf. Fig.~2a), these potential {\sl giants} should have
$Q_{0}>10^{37.5}$ W and be situated in environment with $\rho_{0}<10^{-23}$
kg\,m$^{-3}$.

The above scenerio is not caused by a selection effect. We show that there are
low-luminosity sources which lay in the parts of $\log Q_{0}-\log\rho_{0}$
(Fig.~2a) and $\log E_{\rm tot}-\log Q_{0}$ (Fig.~3b) planes completely avoided
by {\sl giant} sources. They diverge from {\sl giants} and even normal-size
powerful sources by having low jet power $Q_{0}<10^{37.5}$ W and total energy
$E_{\rm tot}<10^{52.5}$ J, thus are not expected to ever reach the giant size.
It is worth to note that some of them already have an age comparable to that of
typical {\sl giants}, in accordance with the model expectations.

\subsection{Acceleration of the cocoon expansion speed}

The statistical analysis in Sect.~5.4 suggests that for $\beta>1.1$ the power
exponent of age $t$ in Eq.(6) is greater than unity, thus the cocoon expansion
along the jet axis may accelerate in time. This effect is also seen in
the time evolution of {\sl fiducial} sources well representing the `clans' of
a few actual sample sources in which each source in the `clan' with almost
identical values of $Q_{0}$ and $\rho_{0}$ is observed at different age (cf.
Sect.~6.2 and Fig.~7a).

Below we argue that the above effect may be real and some arguments come from
the well known hydrodynamical considerations. Beginning from Scheuer
(1974), it is commonly accepted that the ram pressure of the external gas behind
the head of jet is
$\rho_{\rm d}{\cal V}^{2}_{\rm h}\approx Q_{0}/(A_{\rm h}v_{\rm jet})$, where
$\rho_{\rm d}$ is the external density (at the radial distance $d$),
${\cal V}_{\rm h}$ is the head expansion
speed, $A_{\rm h}$ is the head working surface, and the jet bulk velocity
$v_{\rm jet}$ is commonly assumed to be close to the light speed $c$. Therefore,
${\cal V}_{\rm h}$ is a function the root square of the ratio
$Q_{0}/\rho_{\rm d}$ and the reciprocal of a linear size (diameter) of the
working area. $Q_{0}/\rho_{\rm d}$ simply represents an effectiveness of the jet
propagation across the surrounding medium. We can consider this ratio at the
core radius $a_{0}$ and at the source leading head where it is $Q_{0}/\rho_{0}$
and $Q_{0}/\rho_{\rm end}=Q_{0}/\rho_{0}(D/2a_{0})^{-\beta}$, respectively. Both
ratios are derived from the model. Values of the first ratio for {\sl giant}
sources are not much different from those for other sample sources being little
smaller than corresponding values for high-redshift sources, comparable with
those for low-redshift sources, and evidently higher from those for
low-luminosity ones. The second ratio, as a function of $D^{\beta}$, is the
highest for {\sl giant} sources.

The area of $A_{\rm h}$ at a given radius from the AGN centre can be determined
from high-resolution observations of the hotspots. For example, VLBI observations
of selected `compact symmetric objects' (CSO), possibly progenitors of classical
double FRI and/or FRII sources give a linear size of the working area of about
 a few parsecs at a radius of about 50$\sim$100 pc (cf. Owsianik et al. 1998).
Many VLA observations of hotspots in sources of size 10$\sim$20 kpc show
that their sizes are not larger than about 500 pc, while the hotspot sizes in
the lobes of large sources do not exceed  15 kpc. Schoenmakers et al.
(2000) in their study of {\sl giant} radio sources have assumed a typical
hotspot size of 5 kpc at a radius of $0.5\sim 1$ Mpc. After a discovery of the
third largest radio galaxy J1343+3758 (1343+379 in Tables~A1 and A4), Machalski
\& Jamrozy (2000) determined a working surface with diameter of about 13 kpc in
one of its lobes.

From the data in our subsample of {\sl giant} sources we have
$\langle Q_{0}/\rho_{0}\rangle =10^{61.9\pm 0.13}$
W\,m$^{3}$kg$^{-1}$ and $\langle Q_{0}/\rho_{\rm end}\rangle =10^{64.9\pm 0.19}$
W\,m$^{3}$kg$^{-1}$. Dividing the first ratio by $A_{\rm h}\approx 2\,10^{38}$
m$^{2}$ (the working area for hotspot diameter of 0.5 kpc) and the second ratio
by $A_{\rm h}\approx 8\,10^{40}$ m$^{2}$ (for hotspot diameter of 10 kpc), we
obtain ${\cal V}_{\rm h}$ of about 0.12 and 0.19$c$, respectively. Although
these speeds are, at least,
two times higher than those expected from the total linear extent of sources
and their estimated ages (cf. Fig.~1a), their proportion agrees with proportions
found in Sect.~5.4. The above speeds, too high in respect to an upper limit of
${\cal V}_{\rm h}$ (i.e. the quotient $D/2t$), can be caused by either too high
modelled jet power or too low internal density of the core and of the cocoon at
its end. An adjustment of $\beta$ will not work because some increase of
$\beta$ can compensate for too low $\rho_{0}$, but will dramatically lower
$\rho_{\rm end}$. Since the values of $A_{\rm h}$ adopted for this calculation
may be rather overestimated than underestimated, only a decrease of $Q_{0}$ can
fit the observed expansion speeds. Even authors of the KDA paper already noted
that $Q_{0}$ derived from their model is an order of magnitude stronger than
a jet power ($Q^{\rm rs}$) calculated from the values given by Rawlings \&
Saunders (1991).
Indeed, neglecting a thermal gas in the cocoon and taking $\Gamma_{\rm c}=5/3$
and $\beta=3/2$ (as in our calculations), one can found
$Q_{0}/Q^{\rm rs}=1.7\sim 5.5$ depending on the value of ${\cal P}_{\rm hc}$.
Also Lara et al. (2000), assuming
an upper limit temperature of the ambient gas around the high-redshift {\sl
giant} galaxy 0821+695 and deriving the source expansion speed, have determined
its jet power of $(7.1\sim 9)10^{37}$ W which is almost ten times weaker than
our value of $6.9\,10^{38}$ W for that source (cf. Table~A2).
A suspicion that the jet power determined
from the KA model may be too high is supported by much higher than usually
assumed ratios of energy $2Q_{0}t/U_{\rm eq}$.

\subsection{Energy budget}

In a number of studies of {\sl giant} radio sources (e.g. Parma et al. 1996;
Schoenmakers et al. 1998) a fraction
of the total jet energy wasted for adiabatic expansion of the cocoon is usually
assumed to be about 0.5 and used to estimate the jet power $Q_{0}$ if the age
of a source is known (almost always from the spectral ageing analysis). In the
KA model the energy stored in the source (cocoon) is

\[E_{\rm tot}\approx\int\{Q_{0}\,dt-(p_{\rm c}\,d[V_{\rm c}(t)]+
p_{\rm h}\,d[V_{\rm h}(t)])\} \]

\noindent
where $p_{\rm c}\,dV_{\rm c}+p_{\rm h}\,dV_{\rm h}$ is a work done to expand
the cocoon, and $p_{\rm h}$ and $V_{\rm h}$ are the hotspot pressure (cf. Sect.
~4) and its volume, respectively. If $V_{\rm h}$ is neglected, the expansion
work will be $\approx 0.5\,Q_{0}t$; if not, it is dependent on the pressure ratio
${\cal P}_{\rm hc}$  as shown in Sect.~4.2.

In spite of a supposed uncorrectness (possibly a systematic overestimation) of
the jet power derived from the model, the data in our sample fully confirm a
dependence of the energy ratio $2Q_{0}t/U_{\rm eq}$ on the cocoon axial ratio
$AR$, and imply an increase of the fraction of jet energy lost for the adiabatic
slender expansion of the cocoon volume in time. However, the data also suggest
that for a constant $AR$ (a given geometry of cocoon), {\sl giants} tend to have
a smaller ratio of $2Q_{0}t/U_{\rm eq}$ than normal-size sources, i.e. less
energy of the jets was converted into adiabatic expansion of the cocoon. This
may indicate a lower pressure of the external medium surrounding the {\sl giant}
sources than that around shorter ones. That external pressure may be in the
thermal equilibrium with internal pressure in a large part of the cocoon.

\subsection{External pressure of the surrounding medium and its evolution}

A non-relativistic uniform intergalactic medium (IGM) in thermal equilibrium
filling an adiabatically expanding Universe should have an electron pressure
evolving with redshift $p_{\rm IGM}=p^{0}_{\rm IGM}(1+z)^{5}$. The advancing
hotspots of FRII-type radio sources are probably confined by ram pressure of the
IGM. {\sl Giant} sources, with their lobes extended far outside a cluster halo,
have the lowest values of $p_{\rm c}$ and may be useful to determine an upper
limit of $p_{\rm IGM}$. Using a small sample of {\sl giant} sources,
Subrahmanyan \& Saripalli (1993) limited its local value to
 $p^{0}_{\rm IGM}\approx(0.5\sim 2)\,10^{-15}$ N\,m$^{-2}$. A
further study was undertaken by Cotter (1998) who, using a larger sample of 7C
giants with sources out to redshift of about 0.9, confirmed a strong dependence
of the lowest $p_{\rm c}$ and redshift in agreement with a $(1+z)^{5}$ relation.
This observational result has been critically discussed by Schoenmakers et al.
(2000), who have considered possible selection effects in the Cotter's analysis,
and concluded that there was not evidence in their own sample for a cosmological
evolution of $p_{\rm IGM}$. However, they also declare that this hypothesis
cannot be rejected until some low-pressure high-redshift sources are found.

In all above analyses the age of sources was out of consideration. The very
significant correlation between $p_{\rm c}$ and $t$, shown in Sect.~5.7,
strongly suggests that the intrinsic dependence of age on redshift but not the
Malmquist bias is mainly responsible for the apparent correlation between
$p_{\rm c}$ and 1+$z$. Nevertheless, we agree with the Schoenmakers et al.'s
conclusion that until {\sl giant} sources with internal pressures in their lobes
$p_{\rm c}<2\,10^{-15}$ N\,m$^{-2}$ at redshifts of at least $0.6\sim 0.8$ are
not discover, the IGM pressure evolution in the form
$p_{\rm IGM}\propto (1+z)^{5}$ cannot be rejected.

Most {\sl giants} in our sample, but four high-redshift ones, reveal the lowest
pressure in their cocoons. Sharing Subrahmanyan \& Saripalli's arguments,
we can expect those cocoon pressures are
indicative of an upper limit to the present-day external pressure of the IGM,
$p_{\rm IGM}^{0}$. Taking into account the lowest values of $p_{\rm c}$ (cf.
Figs.~3a and 5a), we found $p_{\rm IGM}^{0}<2\,10^{-15}$ N\,m$^{-2}$ in
accordance with their value.
It is worth to emphasize that the above results are obtained from the
analytical model assuming the energy equipartition (cf. Sect.~4.3). This may not
be the case in every part of the source (cocoon). Hardcastle \& Worrall (2000)
estimated gas pressures in the X-ray-emitting medium around normal-size 3CRR
FRII radio galaxies and quasars, and found that, with few exceptions, the
minimum pressures in their lobes, determined under equipartition conditions,
were well below the environment pressures measured from old ROSAT observations.
Therefore, they argued after an additional contribution to the internal pressure
in lobes of those sources including pressure from protons, magnetic fields
differing from their minimum-energy values, or non-uniform filling factors.
Nevertheless, the diffuse lobes of {\sl giants}, extending farther off from a
host galaxy than a typical radius of high-density X-ray-emitting cluster gas,
may be in equilibrium with an ambient medium whose emissivity is not directly
detectable.

\subsection{Dynamical evolution of individual sources}

In Sections 6.3 and 6.4 we have presented the evolutionary tracks of the `fiducial'
sources across the $P$--$D$ and $u_{\rm eq}$--$E_{\rm tot}$ planes, respectively,
calculated according to our modified KDA model, and afterwards compared them
with the predictions of the original KDA model. A predicted evolution of each
`fiducial' source, i.e. its tracks across the above planes, is verified by the
age, luminosity, size, and the energy of a `clan' members (cf. Sect.~6.2).
Therefore, we consider those sources as an individual source observed at
different lifetime epochs, and the relevant `fiducial' source is its
model representation.

Figs.~8 and 9 show that for each of the three `clans' our tracks are steeper than
those resulting from the KDA model, and much better corresponds to the observed 
parameters
of the `clan' sources. Our $P$--$D$ tracks strictly resemble those published by
Blundell et al., as both arise from the dynamical models accounting for not
self-similar evolution of the source cocoon. This is interesting that we get
similar tracks from a different approach, subsequently supporting predictions
of their model. The observed distributions of the
`clan' sources across the $P$--$D$ and $u_{\rm eq}$--$E_{\rm tot}$ planes give
further arguments after a necessity to account for evolving axial ratio of the
cocoon in studies of the dynamical evolution of individual radio sources.

This is worth to emphasize a much higher potential of such analytical models
than it was exploited in previous papers, namely an evolution of the energy
density and total energy stored in a source (its lobes or cocoon) can be
modelled and compared with observations. The tracks of our `clans" across the
$u_{\rm eq}$--$E_{\rm tot}$ plane are steeper than those predicted by the
original KDA model for sources with constant axial ratios. Moreover, the
steepening increases throughout the source lifetime.  This is caused by a
non-linear decline of the adiabatic losses and inflation of the cocoon in time,
as well as a faster decrease of the cocoon pressure in very large sources.
Quantitatively this process is evaluated by the substitution of the evolving
(increasing) value of the pressure ratio ${\cal P}_{\rm hc}(t)$, given by Eq.(7),
into Eq.(3).

\noindent
From our analysis we can conclude as follows:

(1) The data show that the source (cocoon) geometry, described by its axial
ratio, depends on its estimated age.
Incorporation of the cocoon axial ratio evolving with time into the analytical
model of Kaiser et al. (1997) [referred in this paper as KDA], allow us to
reproduce a dynamical evolution of the FRII-type radio sources better than with
their original model which assumes a self-similar expansion of the cocoon of
sources. Therefore we can confirm a conclusion drawn by Blundell et al. (1999)
that throughout the lifetime of an individual source its axial ratio must
steadily increase, thus its expansion cannot be self-similar all the time.
A self-similar expansion seems to be feasible if the power supplied by the jets
is a few orders of magnitude above the minimum-energy values. In other cases
the expansion can only initially be self-similar; a departure from
self-similarity for large and old radio sources is justified by observations of
{\sl giant} sources.

(2) Inserting a statistical correlation between axial ratio and age of the
sample sources into the model, we calculate the evolutionary tracks of a few
`fiducial' sources across the $P$--$D$ and $\rho_{\rm eq}$--$E_{\rm tot}$ planes.
Resultant luminosity, size, and energy density of a given `fiducial' source,
derived for different epochs of its lifetime, are then compared with relevant
parameters of a few real sample sources (called a `clan') having the model
parameters $Q_{0}$ and $\rho_{0}$ similar and common with the fiducial source.
The derived $P$--$D$ tracks appear to be much steeper and better fit the
observational data than those provided with the original KDA model, and are
compatible with those expected from more sophisticated model of Blundell et al.
(1999).

(3) In the case of `clan' sources, we found a slow acceleration of the average 
expansion speed of the cocoon along the jet axis. This effect is predicted with 
the model modified as above, and is caused by a systematic increase of
the ratio between hotspot and cocoon pressures (${\cal P}_{\rm hc}$) with the
age of source and decreasing density of the external environment with
$\beta>1.1$ in its $\beta$-model. The statistics suggests that this acceleration
is also dependent of the jet power.

(4) An evolution of the energetics of sources, predictable from the model, form
another characteristics of their population which can be
constrained by observational data.

(5) {\sl Giant} sources do not form a separate class of radio sources, and do
not reach their extreme sizes exclusively due to some exceptional physical
conditions of the external medium. In average, they are old sources with high
enough jet power evolved in relatively low-density environment. However, for the
observed high-redshift {\sl giants} we found their jets almost ten times
more powerful than those in low-redshift ones, while the core densities of
high-$z$ and low-$z$ {\sl giants} do not differ so much. From the statistical
point of view, we found that for sources with comparable jet power $Q_{0}$, the
size $D$ correlates stronger with their age than anti-correlates with core
density $\rho_{0}$. For {\sl giants} with a comparable $\rho_{0}$ and age, the
size $D$ strongly correlates with $Q_{0}$.

(6) {\sl Giants} possess the lowest equipartition magnetic field strength and
energy density of their cocoons making a difficulty of their detection in
synchrotron emission. However, the accumulated total energy is the highest and
exceeds $3\,10^{52}$ W.

(7) The apparent increase of the lowest internal pressures (observed in the
largest sources) with redshift is mainly caused by the intrinsic dependence of
their age on redshift and dominates over the Malmquist bias suspected by
Schoenmakers et al. (2000) as responsible for this increase. However, a
cosmological  evolution of the IGM cannot be rejected until {\sl giant} sources
with internal pressures in their lobes less than $2\,10^{-15}$ N\,m$^{2}$ at
high redshifts are not discovered.

\section{acknowledgements}

We thank Dr. C. R. Kaiser for explanations of the integration procedures used in
the KDA paper. This work was partly supported by the State Committee for
Scientific Research (KBN) under contract PB 266/PO3/99/17.

\appendix
\section{Observational data and physical parameters of the sample sources}

The observational data for the {\sl giant}-size and normal-size sources are
listed in Table~A1.
Although all columns of Table~A1 are self-explanatory, some more detailed
informations are given below. The overall axial ratio ($AR$) of the sources is
determined from total intensity maps as the ratio of the entire projected source
size to the average of the full deconvolved widths of the two lobes. The later
are measured between $3\sigma$ contours on a radio contour map half-way between
the core and the hot spots or distinct extremities of the source. The reference
to the maps used is given in column 8 of Table~A1. The volume of the sources is
calculated assuming a cylindrical geometry with the diameter equal to the
average width as above. The same geometry has been applied in the dynamical
model used in this paper.

\begin{table*}[t]
\caption{The sample}
\begin{tabular*}{155mm}{@{}lllcrccrl}
\hline
IAU  & Other &$z$& lg$P_{1.4}$ & $D\pm\Delta D$ &AR$\pm\Delta $AR &
lg$V_{\rm c}\pm\Delta$lg$V_{\rm c}$ & Ref. & Spect.\\
name & name  &   & [WHz$^{-1}$sr$^{-1}$] & [kpc] & & [kpc$^{3}$] & map & anal.\\
\hline
GIANTS\\
0109+492   & 3C35     & 0.0670 & 24.53 & 1166$\pm$31 & 3.2$\pm$0.7 & 8.08$\pm$0.17 & 28& 24\\
0136+396   & B2       & 0.2107 & 25.21 & 1555$\pm$35 & 6.0$\pm$1.2 & 7.91$\pm$0.16 & 4 & 6\\
0313+683   & WNB      & 0.0901 & 24.41 & 2005$\pm$34 & 4.2$\pm$0.5 & 8.55$\pm$0.10 & 28& 23\\
0319$-$454 & PKS      & 0.0633 & 24.70 & 2680$\pm$30 & 4.0$\pm$0.8 & 8.98$\pm$0.16 & 22& 22\\
0437$-$244 & MRC      & 0.84   & 26.15 & 1055$\pm$17 & 7.8$\pm$1.5 & 7.18$\pm$0.15 & 5 & 5\\
0813+758   & WNB      & 0.2324 & 25.10 & 2340$\pm$80 & 5.0$\pm$0.5 & 8.60$\pm$0.08 & 25&24\\
0821+695   & 8C       & 0.538  & 25.28 & 2990$\pm$22 & 5.9$\pm$1.0 & 8.77$\pm$0.14 & 7 & 7\\
1003+351   & 3C236    & 0.0988 & 24.76 & 5650$\pm$75 & 9.4$\pm$1.7 & 9.20$\pm$0.14 & 15& 16\\
1025$-$229 & MRC      & 0.309  & 25.28 & 1064$\pm$17 & 5.2$\pm$0.7 & 7.54$\pm$0.11 & 5 & 5\\
1209+745   & 4C74.17  & 0.107  & 24.42 & 1090$\pm$13 & 2.4$\pm$0.5 & 8.25$\pm$0.16 & 2 & 24\\
1232+216   & 3C274.1  & 0.422  & 26.32 & 1024$\pm$15 & 7.4$\pm$1.6 & 7.19$\pm$0.17 & 8 & 1\\
1312+698   & DA340    & 0.106  & 24.76 & 1085$\pm$12 & 4.4$\pm$0.9 & 7.71$\pm$0.16 & 28& 24\\
1343+379   &          & 0.2267 & 24.42 & 3140$\pm$60 & 6.2$\pm$1.1 & 8.80$\pm$0.14 & 28,14& 13\\
1349+647   & 3C292    & 0.71   & 27.02 & 1073$\pm$16 & 6.2$\pm$1.4 & 7.40$\pm$0.18 & 28& 1\\
1358+305   & B2       & 0.206  & 24.86 & 2670$\pm$60 & 3.6$\pm$0.8 & 9.06$\pm$0.17 & 19& 19\\
1543+845   & WNB      & 0.201  & 24.76 & 1950$\pm$25 & 7.6$\pm$1.4 & 8.00$\pm$0.15 & 28& 24\\
1550+202   & 3C326    & 0.0895 & 25.02 & 2510$\pm$55 & 7.0$\pm$1.1 & 8.40$\pm$0.13 & 28& 16\\
2043+749   & 4C74.26  & 0.104  & 24.86 & 1550$\pm$20 & 4.6$\pm$0.8 & 8.14$\pm$0.14 & 28& 24\\
& & &\\
NORMAL\\
0154+286   & 3C55     & 0.720  & 26.79 &  554$\pm$12 & 6.4$\pm$1.5 & 6.51$\pm$0.18 & 9 & 9\\
0229+341   & 3C68.1   & 1.238  & 27.26 &  414$\pm$10 & 4.4$\pm$1.0 & 6.46$\pm$0.18 & 9 & 9\\
0231+313   & 3C68.2   & 1.575  & 27.30 &  190$\pm$4  & 2.8$\pm$0.6 & 5.84$\pm$0.14 & 9 & 9\\
0404+428   & 3C103    & 0.330  & 26.32 &  564$\pm$12 & 6.7$\pm$1.0 & 6.50$\pm$0.12 & 8 & 1\\
0610+260   & 3C154    & 0.5804 & 26.84 &  376$\pm$10 & 2.9$\pm$0.8 & 6.70$\pm$0.21 & 9 & 9\\
0640+233   & 3C165    & 0.296  & 25.94 &  480$\pm$8  & 3.4$\pm$0.8 & 6.88$\pm$0.18 & 8 & 1\\
0642+214   & 3C166    & 0.246  & 26.66 &  187$\pm$15 & 3.1$\pm$0.6 & 5.73$\pm$0.15 & 8 & 1,29\\
0710+118   & 3C175    & 0.768  & 26.85 &  392$\pm$8  & 3.4$\pm$0.9 & 6.61$\pm$0.20 & 9 & 9\\
0828+324   & B2       & 0.0507 & 24.36 &  396$\pm$14 & 3.2$\pm$0.4 & 6.68$\pm$0.10 & 12& 6,20\\
0908+376   & B2       & 0.1047 & 24.39 &  100$\pm$8  & 2.2$\pm$0.2 & 5.21$\pm$0.08 & 18& 20\\
0958+290   & 3C234    & 0.1848 & 25.84 &  460$\pm$8  & 4.6$\pm$1.0 & 6.56$\pm$0.17 & 8 & 1\\
1008+467   & 3C239    & 1.786  & 27.51 &   94$\pm$3  & 2.6$\pm$0.7 & 4.98$\pm$0.21 & 10& 10\\
1012+488   & GB/GB2   & 0.385  & 26.17 &  694$\pm$13 & 2.2$\pm$0.3 & 7.73$\pm$0.11 & 12& 29\\
1030+585   & 3C244.1  & 0.428  & 26.47 &  352$\pm$7  & 5.4$\pm$1.3 & 6.07$\pm$0.19 & 8 & 1\\
1056+432   & 3C247    & 0.749  & 26.85 &  105$\pm$4  & 3.1$\pm$0.7 & 4.98$\pm$0.18 & 10& 10\\
1100+772   & 3C249.1  & 0.311  & 25.94 &  247$\pm$24 & 2.8$\pm$1.0 & 6.18$\pm$0.26 & 9 & 9\\
1111+408   & 3C254    & 0.734  & 26.85 &  107$\pm$4  & 2.5$\pm$0.3 & 5.19$\pm$0.10 & 10& 10\\
1113+295   & B2       & 0.0489 & 24.21 &   97$\pm$5  & 2.2$\pm$0.3 & 5.17$\pm$0.11 & 18& 20\\
1140+223   & 3C263.1  & 0.824  & 27.31 &   45$\pm$5  & 2.2$\pm$0.4 & 4.17$\pm$0.14 & 10& 10\\
1141+354   & GB/GB2   & 1.781  & 26.68 &   97$\pm$3  & 3.0$\pm$0.4 & 4.90$\pm$0.11 & 11& 29\\
1142+318   & 3C265    & 0.8108 & 27.19 &  644$\pm$16 & 5.4$\pm$1.7 & 6.86$\pm$0.24 & 9 & 1,9\\
1143+500   & 3C266    & 1.275  & 27.15 &   37$\pm$2  & 4.0$\pm$0.4 & 3.40$\pm$0.08 & 10& 10\\
1147+130   & 3C267    & 1.144  & 27.17 &  327$\pm$8  & 4.4$\pm$0.7 & 6.15$\pm$0.13 & 9 & 9\\
1157+732   & 3C268.1  & 0.974  & 27.41 &  390$\pm$7  & 4.1$\pm$0.6 & 6.44$\pm$0.12 & 9 & 9\\
1206+439   & 3C268.4  & 1.400  & 27.37 &   87$\pm$4  & 2.8$\pm$0.3 & 4.82$\pm$0.09 & 10& 10\\
1216+507   & GB/GB2   & 0.1995 & 24.93 &  826$\pm$8  & 4.4$\pm$0.6 & 7.36$\pm$0.11 & 12& 29\\
1218+339   & 3C270.1  & 1.519  & 27.48 &  104$\pm$12 & 2.6$\pm$0.5 & 5.12$\pm$0.15 & 10& 10\\
1221+423   & 3C272    & 0.944  & 26.73 &  490$\pm$13 & 3.6$\pm$1.1 & 6.87$\pm$0.23 & 28& 29\\
1241+166   & 3C275.1  & 0.557  & 26.56 &  130$\pm$15 & 2.0$\pm$0.4 & 5.63$\pm$0.16 & 10& 10\\
1254+476   & 3C280    & 0.996  & 27.35 &  110$\pm$13 & 2.4$\pm$0.3 & 5.26$\pm$0.10 & 10& 10\\
1308+277   & 3C284    & 0.2394 & 25.63 &  836$\pm$6  & 6.9$\pm$1.3 & 6.98$\pm$0.15 & 8 & 1\\
1319+428   & 3C285    & 0.0794 & 24.68 &  271$\pm$4  & 2.8$\pm$0.6 & 6.27$\pm$0.17 & 8 & 1\\
1343+500   & 3C289    & 0.967  & 27.02 &   86$\pm$2  & 2.3$\pm$0.2 & 4.97$\pm$0.07 & 10& 10\\
1347+285   & B2       & 0.0724 & 23.60 &   86$\pm$4  & 2.4$\pm$0.3 & 4.94$\pm$0.10 & 18& 20\\
1404+344   & 3C294    & 1.779  & 27.41 &  132$\pm$17 & 3.8$\pm$1.0 & 5.10$\pm$0.20 & 10& 10
\end{tabular*}
\end{table*}
\begin{table*}[t]
\begin{tabular*}{135mm}{@{}lllcrccrl}
1420+198   & 3C300    & 0.270  & 26.00 &  516$\pm$6  & 3.0$\pm$1.2 & 7.08$\pm$0.29 & 8 & 1\\
1441+262   & B2       & 0.0621 & 23.51 &  333$\pm$8  & 4.0$\pm$0.9 & 6.26$\pm$0.18 & 21& 20\\
1522+546   & 3C319    & 0.192  & 25.56 &  390$\pm$15 & 3.2$\pm$0.7 & 6.66$\pm$0.17 & 8 & 1\\
1533+557   & 3C322    & 1.681  & 27.49 &  279$\pm$7  & 3.6$\pm$0.7 & 6.12$\pm$0.15 & 9 & 9\\
1609+660   & 3C330    & 0.549  & 26.93 &  458$\pm$8  & 6.4$\pm$1.2 & 6.27$\pm$0.15 & 9 & 9\\
1609+312   & B2       & 0.0944 & 23.65 &   56$\pm$4  & 2.4$\pm$0.3 & 4.38$\pm$0.10 & 3 & 20\\
1615+324   & 3C332    & 0.1515 & 25.32 &  306$\pm$7  & 4.8$\pm$1.3 & 5.99$\pm$0.21 & 3 & 20\\
1618+177   & 3C334    & 0.555  & 26.45 &  430$\pm$15 & 2.8$\pm$0.4 & 6.90$\pm$0.12 & 9 & 9\\
1658+302   & B2       & 0.0351 & 23.39 &  120$\pm$10 & 2.2$\pm$0.2 & 5.45$\pm$0.07 & 21& 20\\
1723+510   & 3C356    & 1.079  & 26.96 &  643$\pm$13 & 7.9$\pm$1.0 & 6.52$\pm$0.10 & 9 & 9\\
1726+318   & 3C357    & 0.1664 & 25.43 &  395$\pm$10 & 3.0$\pm$0.6 & 6.73$\pm$0.16 &21,3&20\\
1957+405   & CygA     & 0.0564 & 27.25 &  185$\pm$3  & 3.8$\pm$0.5 & 5.54$\pm$0.11 & 9 & 9\\
2019+098   & 3C411    & 0.467  & 27.05 &  201$\pm$5  & 2.6$\pm$0.6 & 5.97$\pm$0.18 & 26& 26\\
2104+763   & 3C427.1  & 0.572  & 26.75 &  173$\pm$5  & 2.9$\pm$0.7 & 5.68$\pm$0.19 & 9 & 9\\
\hline
{\bf References}\\
\end{tabular*}
\vspace{2mm}
\begin{tabular*}{135mm}{@{}rlrlrl}
(1)& Alexander \& Leahy 1987&         (11)& Machalski \& Condon 1983&   (21)& de Ruiter et al. 1986\\
(2)& van Breugel \& Willis 1981&      (12)& Machalski \& Condon 1985&   (22)& Saripalli et al. 1994\\
(3)& Fanti et al. 1986&               (13)& Machalski \& Jamrozy 2000&  (23)& Schoenmakers et al. 1998\\
(4)& Hine 1979&                       (14)& Machalski et al. 2001&      (24)& Schoenmakers et al. 2000\\
(5)& Ishwara-Chandra \& Saikia 1999&  (15)& Mack et al. 1997&           (25)& Schoenmakers et al. 2001\\
(6)& Klein et al. 1995&               (16)& Mack et al. 1998&           (26)& Spangler \& Pogge 1984\\
(7)& Lara et al. 2000&                (17)& Myers \& Spangler 1985&     (27)& FIRST (Becker et al. 1996)\\
(8)& Leahy \& Williams 1984&          (18)& Parma et al. 1986&          (28)& NVSS (Condon et al. 1998)\\
(9)& Leahy et al. 1989&               (19)& Parma et al. 1996&          (29)& this paper\\
(10)& Liu et al. 1992&                (20)& Parma et al. 2000
\end{tabular*}
\end{table*}

Table~A2 contains the physical parameters of the sample sources derived from the
model. The entries are:

{\sl Columns 2, 3 and 4:} Age, equipartition energy density, and equipartition
magnetic field strength with their estimated uncertainty, respectively -- as
described in Sect.~3.

{\sl Columns 5, 6 and 7:} Logarithms of the jet power, initial density of the
external medium at the core radius $a_{0}$, and the cocoon pressure,
respectively, derived from the model.

{\sl Column 8:} Logarithms of the total source (cocoon) energy.

{\sl Column 9:} Ratio between the total energy of twin jets and source energy.

\begin{table*}[t]
\caption{Age and physical parameters}
\begin{tabular*}{165mm}{@{}lrrlllrcl}
\hline
Source & t & lg$u_{\rm eq}\pm\Delta\lg u_{\rm eq}$ & $B_{\rm eq}\pm\Delta B_{\rm eq}$ & lg$Q_{0}$ & lg$\rho_{0}$ &
                                        lg$p_{\rm c}$ & lg$U_{\rm eq}\pm\Delta \lg U_{\rm eq}$ &\underline{$2Q_{0}t$}\\
       &[Myr]&[Jm$^{-3}$] & [nT] &  [W] & [kgm$^{-3}$] & [Nm$^{-2}$] & [J] &$U_{\rm eq}$\\
\hline
GIANTS\\
0109+492   &  96$\pm$18 & $-13.74\pm 0.12$ & 0.14$\pm$0.02 &  37.69 & $-$24.03 & $-$13.86 & 52.81$\pm$0.28 &4.6$\pm$2.0\\
0136+396   &  89$\pm$17 & $-13.24\pm 0.17$ & 0.25$\pm$0.05 &  38.29 & $-$23.30 & $-$13.36 & 53.14$\pm$0.31 &8.0$\pm$4.2\\
0313+683   & 140$\pm$24 & $-14.11\pm 0.14$ & 0.09$\pm$0.015&  37.82 & $-$23.87 & $-$14.11 & 52.91$\pm$0.23 &7.2$\pm$2.6\\
0319$-$454 & 180$\pm$40 & $-13.83\pm 0.13$ & 0.13$\pm$0.02 &  38.07 & $-$23.79 & $-$14.15 & 53.62$\pm$0.28 &3.2$\pm$1.3\\
0437$-$244 &  19$\pm$6  & $-12.37\pm 0.15$ & 0.68$\pm$0.12 &  39.20 & $-$23.37 & $-$12.46 & 53.28$\pm$0.29&10.0$\pm$3.5\\
0813+758   &  84$\pm$4  & $-13.60\pm 0.15$ & 0.17$\pm$0.03 &  38.47 & $-$23.90 & $-$13.79 & 53.47$\pm$0.22 &5.3$\pm$2.5\\
0821+695   &  84$\pm$10 & $-13.37\pm 0.17$ & 0.21$\pm$0.04 &  38.84 & $-$23.68 & $-$13.64 & 53.87$\pm$0.30 &5.0$\pm$2.8\\
1003+351   & 127$\pm$18 & $-14.25\pm 0.14$ & 0.08$\pm$0.013&  38.62 & $-$23.83 & $-$14.34 & 53.42$\pm$0.27&12.8$\pm$6.1\\
1025$-$229 &  64$\pm$12 & $-12.82\pm 0.22$ & 0.40$\pm$0.10 &  38.23 & $-$23.25 & $-$13.10 & 53.19$\pm$0.31 &4.4$\pm$2.3\\
1209+745   & 110$\pm$20 & $-13.79\pm 0.13$ & 0.13$\pm$0.02 &  37.60 & $-$24.23 & $-$13.99 & 52.93$\pm$0.28 &5.3$\pm$1.5\\
1232+216   &  22$\pm$3  & $-12.29\pm 0.32$ & 0.74$\pm$0.27 &  39.20 & $-$23.20 & $-$12.38 & 53.37$\pm$0.43 &9.4$\pm$8.0\\
1312+698   &  55$\pm$5  & $-13.44\pm 0.14$ & 0.19$\pm$0.03 &  37.95 & $-$23.97 & $-$13.56 & 52.74$\pm$0.29 &5.7$\pm$3.2\\
1343+379   &  94$\pm$16 & $-14.20\pm 0.33$ & 0.08$\pm$0.03 &  38.24 & $-$24.16 & $-$14.25 & 53.07$\pm$0.41 &8.8$\pm$6.8\\
1349+647   &  16$\pm$4  & $-11.84\pm 0.17$ & 1.25$\pm$0.25 &  39.79 & $-$23.30 & $-$12.07 & 54.03$\pm$0.33 &5.8$\pm$3.0\\
1358+305   & 125$\pm$25 & $-13.95\pm 0.14$ & 0.12$\pm$0.02 &  38.34 & $-$24.12 & $-$14.10 & 53.58$\pm$0.30 &4.6$\pm$2.2\\
1543+845   & 130$\pm$21 & $-13.46\pm 0.23$ & 0.19$\pm$0.05 &  38.03 & $-$23.00 & $-$13.61 & 53.01$\pm$0.35 &8.6$\pm$5.6\\
1550+202   & 134$\pm$27 & $-13.63\pm 0.27$ & 0.16$\pm$0.05 &  38.26 & $-$23.21 & $-$13.73 & 53.24$\pm$0.36 &8.9$\pm$5.7\\
2043+749   &  64$\pm$11 & $-13.57\pm 0.10$ & 0.17$\pm$0.02 &  38.14 & $-$24.06 & $-$13.74 & 53.04$\pm$0.23 &5.1$\pm$1.9\\
& & &\\
NORMAL\\
0154+286   & 13.0$\pm$1.7& $-11.55\pm 0.16$ & 1.74$\pm$0.32 &  39.43 & $-$22.98 & $-$11.72 & 53.43$\pm$0.32 &8.2$\pm$5.0\\
0229+341   & 11.3$\pm$1.8& $-11.31\pm 0.19$ & 2.31$\pm$0.43 &  39.69 & $-$22.91 & $-$11.33 & 53.62$\pm$0.35 &8.4$\pm$5.4\\
0231+313   &  4.6$\pm$0.3& $-10.71\pm 0.16$ & 4.58$\pm$0.85 &  39.64 & $-$23.55 & $-$11.03 & 53.60$\pm$0.29 &3.2$\pm$1.9\\
0404+428   & 17.7$\pm$2.4& $-11.91\pm 0.17$ & 1.15$\pm$0.23 &  39.02 & $-$22.96 & $-$12.01 & 53.06$\pm$0.30&10.2$\pm$5.5\\
0610+260   & 45.2$\pm$3.8& $-11.72\pm 0.18$ & 1.43$\pm$0.30 &  39.00 & $-$22.20 & $-$11.54 & 53.45$\pm$0.36&10.2$\pm$7.6\\
0640+233   & 60$\pm$10   & $-12.34\pm 0.16$ & 0.70$\pm$0.13 &  38.33 & $-$22.66 & $-$12.30 & 53.01$\pm$0.32 &7.9$\pm$4.5\\
0642+214   & 14.0$\pm$5.0& $-11.71\pm 0.22$ & 1.44$\pm$0.37 &  38.87 & $-$22.71 & $-$11.23 & 52.56$\pm$0.34&17.9$\pm$7.9\\
0710+118   & 34.5$\pm$4.5& $-11.54\pm 0.16$ & 1.76$\pm$0.32 &  39.09 & $-$22.31 & $-$11.52 & 53.54$\pm$0.34 &7.8$\pm$5.0\\
0828+324   & 59$\pm$9    & $-13.27\pm 0.18$ & 0.24$\pm$0.05 &  37.20 & $-$23.60 & $-$13.24 & 51.88$\pm$0.27 &7.8$\pm$3.6\\
0908+376   & 28.4$\pm$4.4& $-12.33\pm 0.16$ & 0.71$\pm$0.13 &  36.83 & $-$23.36 & $-$12.39 & 51.35$\pm$0.23 &5.5$\pm$2.0\\
0958+290   & 22.0$\pm$4.0& $-12.11\pm 0.15$ & 0.92$\pm$0.16 &  38.53 & $-$23.31 & $-$12.32 & 52.92$\pm$0.30 &5.7$\pm$2.9\\
1008+467   &  2.8$\pm$0.2& $-10.28\pm 0.16$ & 7.48$\pm$1.39 &  39.66 & $-$23.21 & $-$10.36 & 53.17$\pm$0.35 &5.5$\pm$4.0\\
1012+488   & 38.5$\pm$5.5& $-12.83\pm 0.13$ & 0.40$\pm$0.06 &  38.81 & $-$23.93 & $-$12.80 & 53.37$\pm$0.23 &6.7$\pm$2.6\\
1030+585   & 14.0$\pm$2.0& $-11.56\pm 0.16$ & 1.72$\pm$0.32 &  38.98 & $-$22.84 & $-$11.62 & 52.98$\pm$0.33 &8.9$\pm$5.5\\
1056+432   &  3.2$\pm$0.3& $-10.72\pm 0.14$ & 4.55$\pm$0.71 &  39.18 & $-$23.44 & $-$10.81 & 52.73$\pm$0.31 &5.7$\pm$3.5\\
1100+772   & 31.5$\pm$5.5& $-11.92\pm 0.16$ & 1.14$\pm$0.21 &  38.24 & $-$22.87 & $-$11.95 & 52.73$\pm$0.38 &6.5$\pm$4.6\\
1111+408   &  3.1$\pm$0.2& $-10.75\pm 0.15$ & 4.40$\pm$0.76 &  39.20 & $-$23.79 & $-$10.97 & 52.91$\pm$0.24 &3.8$\pm$1.9\\
1113+295   & 18.8$\pm$2.7& $-12.42\pm 0.16$ & 0.64$\pm$0.12 &  36.84 & $-$23.84 & $-$12.51 & 51.22$\pm$0.26 &5.0$\pm$2.2\\
1140+223   &  1.7$\pm$0.6& $-9.87 \pm 0.11$ & 12.0$\pm$1.50 &  39.38 & $-$23.27 & $-$10.02 & 52.77$\pm$0.24 &4.4$\pm$1.0\\
1141+354   &  3.4$\pm$0.8& $-10.69\pm 0.16$ & 4.68$\pm$0.87 &  39.03 & $-$23.44 & $-$10.85 & 52.68$\pm$0.26 &4.8$\pm$2.4\\
1142+318   & 22.0$\pm$5.0& $-11.66\pm 0.16$ & 1.54$\pm$0.29 &  39.64 & $-$22.52 & $-$11.57 & 53.67$\pm$0.37&13.0$\pm$8.1\\
1143+500   &  1.2$\pm$0.5&  $-9.67\pm 0.16$ & 15.1$\pm$2.80 &  39.32 & $-$22.65 &  $-$9.58 & 52.20$\pm$0.23&10.0$\pm$4.2\\
1147+130   & 12.5$\pm$3.6& $-11.10\pm 0.14$ & 2.93$\pm$0.48 &  39.51 & $-$22.60 & $-$11.16 & 53.52$\pm$0.26 &7.7$\pm$2.4\\
1157+732   & 12.6$\pm$2.6& $-11.17\pm 0.16$ & 2.70$\pm$0.49 &  39.73 & $-$22.73 & $-$11.20 & 53.74$\pm$0.27 &7.8$\pm$3.2\\
1206+439   &  3.0$\pm$0.3& $-10.34\pm 0.12$ & 7.00$\pm$1.00 &  39.52 & $-$23.04 & $-$10.32 & 52.95$\pm$0.20 &7.1$\pm$2.6\\
1216+507   & 50$\pm$15   & $-13.11\pm 0.18$ & 0.29$\pm$0.06 &  37.98 & $-$23.73 & $-$13.29 & 52.72$\pm$0.28 &5.8$\pm$1.9\\
1218+339   &  3.9$\pm$1.1& $-10.35\pm 0.17$ & 6.90$\pm$1.35 &  39.60 & $-$23.00 & $-$10.41 & 53.24$\pm$0.30 &5.7$\pm$2.4\\
1221+423   & 18.0$\pm$8.0& $-11.94\pm 0.16$ & 1.12$\pm$0.21 &  39.25 & $-$23.26 & $-$11.90 & 53.40$\pm$0.36 &8.1$\pm$3.1\\
1241+166   &  4.2$\pm$0.4& $-11.40\pm 0.16$ & 2.07$\pm$0.38 &  39.01 & $-$24.21 & $-$11.44 & 52.70$\pm$0.30 &5.5$\pm$3.3\\
1254+476   &  2.8$\pm$0.3& $-10.68\pm 0.16$ & 4.77$\pm$0.88 &  39.59 & $-$23.64 & $-$10.69 & 53.05$\pm$0.25 &6.2$\pm$2.9\\
1308+277   & 60$\pm$9    & $-12.60\pm 0.18$ & 0.52$\pm$0.11 &  38.37 & $-$22.59 & $-$12.63 & 52.85$\pm$0.31&12.6$\pm$7.2\\
1319+428   & 110$\pm$20  & $-12.74\pm 0.16$ & 0.43$\pm$0.08 &  37.15 & $-$22.79 & $-$12.90 & 52.03$\pm$0.30 &9.1$\pm$4.9\\
1343+500   &  4.0$\pm$0.5& $-10.71\pm 0.16$ & 4.61$\pm$0.85 &  39.17 & $-$23.28 & $-$10.67 & 52.73$\pm$0.22 &7.0$\pm$2.7\\
1347+285   & 21.4$\pm$4.2& $-12.31\pm 0.13$ & 0.73$\pm$0.11 &  36.35 & $-$23.85 & $-$12.72 & 51.10$\pm$0.22 &2.4$\pm$0.8\\
1404+344   &  2.8$\pm$0.2& $-10.59\pm 0.16$ & 5.23$\pm$0.97 &  39.72 & $-$23.15 & $-$10.50 & 52.98$\pm$0.34 &9.8$\pm$6.7\\
\end{tabular*}
\end{table*}
\begin{table*}[t]
\begin{tabular*}{153mm}{@{}lrrlllrcl}
1420+198   & 43$\pm$6    & $-12.27\pm 0.14$ & 0.76$\pm$0.12 &  38.49 & $-$23.22 & $-$12.47 & 53.28$\pm$0.39 &4.4$\pm$3.3\\
1441+262   & 77$\pm$13   & $-13.30\pm 0.19$ & 0.23$\pm$0.05 &  36.49 & $-$23.40 & $-$13.46 & 51.43$\pm$0.35 &5.6$\pm$3.5\\
1522+546   & 43$\pm$7    & $-12.33\pm 0.15$ & 0.71$\pm$0.12 &  38.07 & $-$23.13 & $-$12.49 & 52.80$\pm$0.30 &5.1$\pm$2.7\\
1533+557   &  7.3$\pm$0.6& $-11.00\pm 0.16$ & 3.28$\pm$0.60 &  39.82 & $-$23.01 & $-$10.99 & 53.59$\pm$0.30 &7.9$\pm$4.7\\
1609+660   & 20.3$\pm$3.3& $-11.39\pm 0.18$ & 2.10$\pm$0.43 &  39.35 & $-$22.19 & $-$11.36 & 53.35$\pm$0.31&12.9$\pm$7.1\\
1609+312   & 11.5$\pm$2.5& $-12.08\pm 0.14$ & 0.95$\pm$0.15 &  36.42 & $-$24.00 & $-$12.39 & 50.77$\pm$0.23 &3.3$\pm$1.0\\
1615+324   & 47$\pm$9    & $-12.24\pm 0.15$ & 0.79$\pm$0.14 &  37.78 & $-$22.40 & $-$12.18 & 52.22$\pm$0.34&10.8$\pm$6.3\\
1618+177   & 31.6$\pm$3.7& $-11.91\pm 0.15$ & 1.15$\pm$0.20 &  38.83 & $-$23.10 & $-$12.07 & 53.46$\pm$0.26 &4.7$\pm$2.2\\
1658+302   & 48.5$\pm$8.5& $-13.05\pm 0.17$ & 0.31$\pm$0.06 &  36.10 & $-$23.67 & $-$13.12 & 50.87$\pm$0.20 &5.2$\pm$1.5\\
1723+510   & 20.0$\pm$2.6& $-11.54\pm 0.18$ & 1.76$\pm$0.36 &  39.55 & $-$22.29 & $-$11.53 & 53.45$\pm$0.27&16.0$\pm$7.7\\
1726+318   & 27.0$\pm$5.0& $-12.48\pm 0.14$ & 0.60$\pm$0.10 &  38.14 & $-$23.76 & $-$12.67 & 52.72$\pm$0.29 &4.5$\pm$2.1\\
1957+405   & 12.6$\pm$3.0& $-10.39\pm 0.15$ & 6.60$\pm$1.18 &  39.39 & $-$22.03 & $-$10.61 & 53.62$\pm$0.25 &4.7$\pm$1.6\\
2019+098   &  9.2$\pm$1.2& $-11.14\pm 0.10$ & 2.80$\pm$0.32 &  39.28 & $-$23.20 & $-$11.22 & 53.30$\pm$0.27 &5.6$\pm$2.7\\
2104+763   &  7.7$\pm$0.6& $-10.87\pm 0.14$ & 3.80$\pm$0.63 &  39.06 & $-$23.28 & $-$11.25 & 53.28$\pm$0.31 &2.9$\pm$1.9\\
\hline
\end{tabular*}
\end{table*}

\end{document}